\documentclass[10pt]{iopart} 

\usepackage[english]{babel}

\usepackage{iopams} 

\expandafter\let\csname equation*\endcsname\relax
\expandafter\let\csname endequation*\endcsname\relax
\usepackage{amsmath}
\usepackage{amssymb}
\usepackage{color}
\usepackage{geometry}

\usepackage{ifthen} 
\usepackage{ifpdf}
\ifpdf
	\usepackage[pdftex]{graphicx}
	\usepackage[pdftex]{hyperref} 
	\hypersetup{colorlinks=true,
		linkcolor=blue,
		citecolor=blue,
		urlcolor=blue,
	  pdfauthor={Paul HEINRICH paul.heinrich@ipp.mpg.de},
	  pdfsubject={2023 P. Heinrich ASDEX Runaway waves paper},
	  pdftitle={2023 P. Heinrich ASDEX Runaway waves paper}
	}
	\DeclareGraphicsExtensions{.pdf}
\else
	\usepackage[dvips]{graphicx}
	\usepackage{epsfig}
	\DeclareGraphicsExtensions{.eps}
	\usepackage{url}
\fi

\usepackage[numbers,sort&compress]{natbib}
\usepackage{hypernat}

\usepackage{bmpsize}

\usepackage{caption}
\usepackage{mwe}

\usepackage{subfigure}
\usepackage{lipsum}
\usepackage{longtable}
\usepackage{hhline}
\usepackage{ulem}

\usepackage{tikz}

\usepackage{makecell}

\newcommand{\habb}[1]{{\color{black}\textnormal{#1}}}

\newcommand{\highlight}[1]{\textbf{#1}}

\newcommand*\circled[1]{\tikz[baseline=(char.base)]{%
            \node[shape=circle,draw,inner sep=1.5pt] (char) {#1};}}
\newcommand*\circledtext[1]{\tikz[baseline=(char.base)]{%
            \node[shape=circle,draw,inner sep=0.8] (char) {#1};}}

\usepackage{enumitem}
\usepackage[figuresleft]{rotating}

\begin{document}
\hyphenation{ASDEX}

\title{Characteristics of the Alfv\'enic activity during the current quench in ASDEX Upgrade}

\author{P.~Heinrich$^{1, 2}$, G.~Papp$^{1}$, Ph.~Lauber$^{1}$, G.~Pautasso$^{1}$, M.~Dunne$^{1}$, M.~Maraschek$^{1}$, V.~Igochine$^{1}$, O.~Linder$^{1}$, the ASDEX Upgrade Team$^a$, and the EUROfusion Tokamak Exploitation Team$^b$}
\address{$^1$Max Planck Institute for Plasma Physics, Boltzmannstr. 2, D-85748 Garching, Germany}
\address{$^2$Physics department of TUM, James-Franck-Str. 1, 85748 Garching, Germany}
\address{$^a$See the author list of \href{https://doi.org/10.1088/1741-4326/ad249d}{H.~Zohm et~al. 2024 Nucl. Fusion}} 
\address{$^b$See the author list of ''Progress on an exhaust solution for a reactor using EUROfusion multi-machines capabilities'' by E. Joffrin et al. to be published in Nuclear Fusion Special Issue: Overview and Summary Papers from the 29th Fusion Energy Conference (London, UK, 16-21 October 2023)} 

\date{\today}

\begin{abstract}
    ASDEX Upgrade has developed multiple massive gas injection~(MGI) scenarios to investigate runaway electron~(RE) dynamics. 
    During the current quench of the MGI induced disruptions, Alfv\'enic activity is observed in the 300--800~kHz range. With the help of a mode tracing algorithm based on Fourier spectrograms, mode behaviour was classified for 180 discharges. The modes have been identified as global Alfv\'en eigenmodes using linear gyrokinetic MHD simulations. Changes in the Alfv\'en continuum during the quench are proposed as explanation for the strong frequency sweep observed. A systematic statistical analysis shows no significant connection of the mode characteristics to the dynamics of the subsequent runaway electron beams. In our studies, the appearance and amplitude of the modes does not seem to affect the potential subsequent runaway beam. Beyond the scope of the 180 investigated dedicated RE experiments, the Alfv\'enic activity is also observed in natural disruptions with no RE beam forming. 
\end{abstract}

\maketitle
\ioptwocol 

\section{Introduction \label{sec:Introduction}} 
Widely used techniques for disruption mitigation in tokamaks include massive gas injection~(MGI)~\cite{Pautasso2015Assimilation, Pautasso_2016, Pautasso_2020} and shattered pellet injection~(SPI)~\cite{Jachmich_2022, Sheikh_2021}. Hereby, a large amount of material is injected into the plasma as gas puff or as frozen pellet fragments, respectively.
Even though there has been significant progress on the physics basis and design of the ITER disruption mitigation system~(DMS)~\cite{Lehnen_2023_FEC, Jachmich_2022, Sheikh_2021, Gebhart2021, Jachmich_2nd_IAEA_disruptions_2022}, effective disruption mitigation for the next generation of fusion devices such as ITER or DEMO remains a challenging task~\cite{Nardon_2021, vallhagen_2023, pusztai_2023}. In particular, developing a robust runaway electron~(RE) mitigation technique for an activated wall phase with nuclear primary runaway electron sources has been proven difficult~\cite{Boozer_2017,Vallhagen_2022}. A beam of runaway electrons with multi-MeV energies rises due to high toroidal electric fields induced during a disruption~\cite{Breizman_2019}. Even a small seed of REs can self-amplify due to collisions between the fast seed electrons and the bulk, leading to an avalanche effect~\cite{Rosenbluth_1997, Hesslow_2019}. Ideally, disruption mitigation scenarios aim to avoid the formation of high-current RE beams, which can cause severe damage to the machine~\cite{Matthews_2016}.

The benign termination of the RE beam~\cite{Paz-Soldan_2021, Reux_2022, Sheikh_2024} is a promising candidate for a mitigation technique.
Plasma waves might contribute to the avoidance of the formation of strong RE beams, as the level of magnetic perturbations~\cite{Svensson_2021} and Alfv\'enic activity in post-disruption plasmas~\cite{Lier_2021,Lier_2023} can have a strong influence on a subsequent runaway current formation. In the contributions by Lvovskiy \textit{et al.}~\cite{Lvovskiy_2018, Lvovskiy_2023}, they present instabilities observed during disruptions in the tokamak \mbox{DIII-D}, which could deconfine the initial seed electrons, hence suppressing the formation of a strong RE beam. 

In this paper we present a detailed investigation on post-disruption modes observed in the tokamak ASDEX Upgrade~(AUG). We discuss mode characteristics, different statistical analysis techniques as well as simulation results. The modes described in this paper were identified as global Alfv\'en eigenmodes~(GAEs). The paper is organised as follows. In section~\ref{sec:data_base} the setup and discharge parameters of the experiments are described. The tracing algorithm used for the analysis is detailed in section~\ref{sec:MT}. The observations and results of the mode tracing are discussed in section~\ref{sec:observations}. Section~\ref{sec:Classification} presents the mode classification results. Section~\ref{sec:statistical_analysis} covers the statistical analysis of the modes. Simulation analysis of the mode characteristics is presented in section~\ref{sec:theory}.

\section{Experimental dataset \label{sec:data_base}} 

The database analysed in this paper consists of 180 ASDEX Upgrade discharges conducted with the primary goal of runaway electron generation. It ranges from shot number \#31318~(July 2014) to \#38088~(July 2020). The discharges are all inner wall limited, circular plasmas~(elongation \(\kappa \approx 1.1\))~\cite{Pautasso_2020}. A summary of global parameters for these discharges is shown in table~\ref{tab:global parameters}.

\begin{table}[h]
	\caption{Global parameter ranges for the discharges in the database. The numbers inside the parentheses represent the number of discharges within this parameter range, with a total of 180.}
        \label{tab:global parameters}
        \centering
        \def\arraystretch{1.5} 
        \begin{tikzpicture}
            \node (table) [inner sep=0pt] {
            \begin{tabular}{cl l l}
                \multicolumn{3}{c}{GLOBAL PARAMETER RANGES} \\
                \hline
                \multicolumn{3}{c}{Toroidal magnetic field \(B_\textrm{t}\) [T]} \\
                \hline
                $<$ 2.45 (17) &  2.45 -- 2.55 (151) & 2.7 -- 2.9 (12)\\
                \hline
                \hline
                \multicolumn{3}{c}{Plasma current \(I_\textrm{p}\) [MA]} \\
                \hline
                $<$ 0.75 (15) & 0.75 -- 0.85 (149) & 0.85 -- 1.17 (16)\\
                \hline
                \hline
                \multicolumn{3}{c}{Pre disruption density \(n_\textrm{pre}\) \(\left[\textrm{m}^{-3}\right]\)} \\
                \hline
                \multicolumn{3}{c}{around the target density of \(3\cdot10^{19}\ \textrm{m}^{-3}\)}\\
                \hline
                \hline
                \multicolumn{3}{c}{Edge safety factor \(q_\textrm{95}\)} \\
                \hline
                $<$ 3.7 (27) & 3.7 -- 4.05 (151) & 4.22 \& 6.01 (2) \\
                \hline
                \hline
                \multicolumn{3}{c}{ECRH power \(\textrm{P}_\textrm{ECRH}\) [MW]} \\
                \hline
                \multicolumn{3}{c}{\begin{tabular}{l l l l} 0 (40) & $<$ 2.0 (63) & 2.0 -- 2.6 (100) & $>$ 2.6 (17)\end{tabular}}\\
                \hline
                \hline
                \multicolumn{3}{c}{MGI gas type} \\
                \hline
                \multicolumn{3}{c}{\begin{tabular}{l l l l} Ar (166) & Ne (3) & Kr (5) & Ar + \(\textrm{D}_2\) (6)\end{tabular}}
            \end{tabular}
            };
            \draw [rounded corners=.5em] (table.north west) rectangle (table.south east);
        \end{tikzpicture}
        
    \end{table}
    
\begin{figure*}
	\centering
	\subfigure[Typical RE scenario with trigger of the MGI valve at t~=~1.0~s.]{\includegraphics[width=0.49\linewidth]{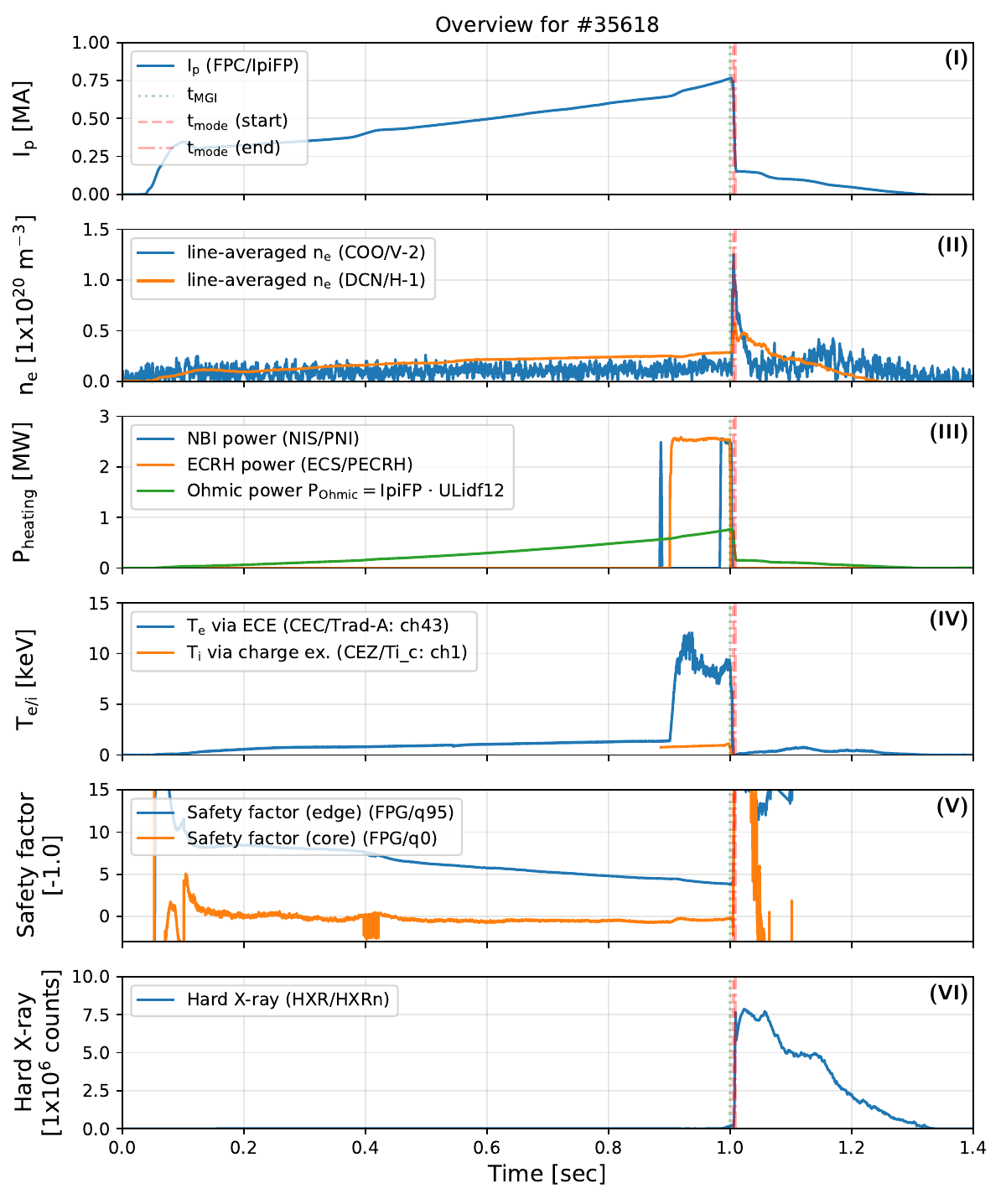}\label{fig:scenario_35618_entire}}
	\hfill
	\subfigure[Zoom of (a) with mode activity time window highlighted.]{\includegraphics[width=0.49\linewidth]{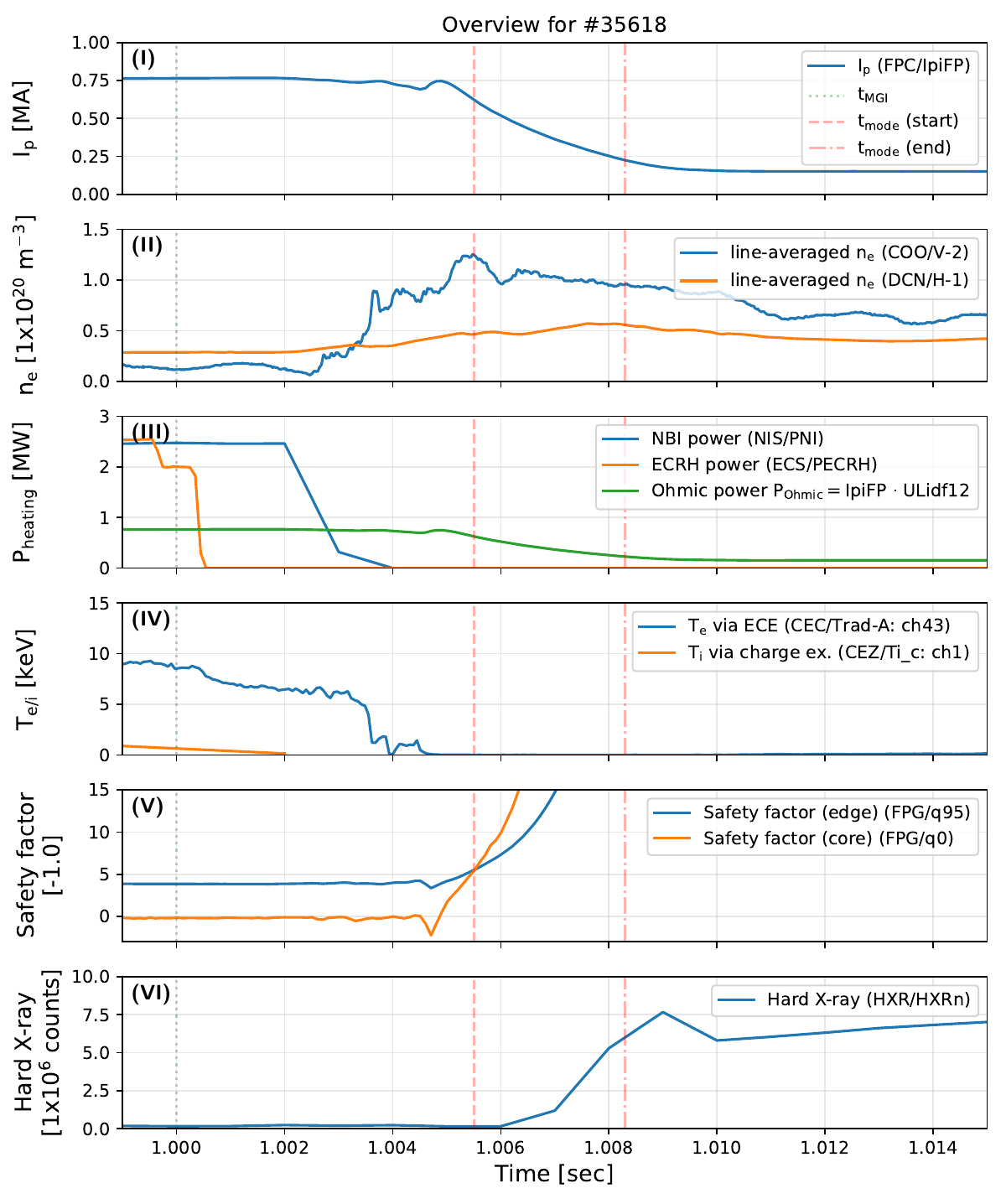}\label{fig:scenario_35618_zoomed}}
    	\caption{Reference discharge \#35618 as a representative shot for the discharges within the database. The plasma current in~(I), line-averaged electron density from interferometry~(II), heating powers~(III), electron (core Electron Cyclotron Emission -- ECE) and ion temperatures (CXRS)~(IV), edge/core safety factors~(V), and hard X-ray count~(VI) are given. In the zoomed-in version in~(b), the time window of the mode for this discharge is also highlighted.}
    	\label{fig:scenarios}
\end{figure*}

In figure~\ref{fig:scenarios} the temporal evolution of the reference discharge of this paper (\#35618) is indicated. We divide the discharges into three phases. The first phase is the current ramp-up, leading to the pre-timed massive gas injection. During the first second of the discharge the plasma current~(\(I_\textnormal{p}\)) is ramped up to values typically around 800~kA and the target density (pre-disruption plasma density~\(n_\textnormal{pre}\) or electron density~\(n_\textnormal{e}\)) is around \(3\cdot 10^{19}\ \textrm{m}^{-3}\). At around \(t = 1.00\ \textrm{s}\), an in-vessel massive gas injection~\cite{Pautasso2015Assimilation, Pautasso_2016, Pautasso_2020} is triggered, which leads to the intended disruption. Most of the cases feature argon injection at different gas pressures, with a minority of the shots testing neon or krypton injection. The discharges typically do not reach flat-top and are disrupted during the ramp-up phase. This is motivated by two reasons: to increase the chance of runaway generation via the presence of the ramp-up loop voltage, and to preempt the potential formation of locked modes due to the relatively low density operation. The exact injection timing slightly varied for 20 of the 180 discharges inside the database. The reason for this variation is a scan of pre-disruption plasma current. Disrupting the ramp-up at different times leads to runaway formation from different starting currents, while keeping the same loop voltage during the ramp-up and at the time of the injection. Core Electron Cyclotron Resonance Heating (ECRH) heating is applied in some cases 100~ms before the timed MGI to facilitate RE generation and to provide a core temperature scan. Additionally, a Neutral Beam Injection (NBI) blip was applied to enable the measurements of beam diagnostics like Charge Exchange Recombination Spectroscopy (CXRS) (see ion temperature \(T_\textnormal{i}\) measurement in figure~\ref{fig:scenarios}). 
    
To achieve reliable runaway beam formation, most of the time argon is used as MGI gas -- in this database 166 out of 180 discharges (see table~\ref{tab:global parameters}). Argon quantities injected vary between \(3\cdot 10^{20}\) and \(4.8 \cdot 10^{21}\) atoms. In the second phase, the injected gas penetrates the plasma and starts to radiate away the previously confined energy~\cite{Linder_2020}. Within a few milliseconds the plasma cools down from around 10~keV of core temperature to a few~eV, followed by the current quench, discussed in more detail in section~\ref{sec:observations}.

As the experiments were designed to investigate runaway electron generation~\cite{Pautasso_2016, Pautasso_2020}, the last phase of the discharges is mostly characterised by the presence of a RE plateau. Around 77\% of the discharges in the set show a runaway current above 50~kA. For plasma currents below this value plasma position control is difficult and the shot typically ends in a VDE. A more detailed description of the scenario and these experiments is given in the papers by Pautasso \textit{et al.}~\cite{Pautasso_2016, Pautasso_2020}.
 
Of the 180 discharges in this dataset, virtually all of them have shown signatures of Alfv\'enic frequency modes in the current quench. The magnetic pick-up coils (Mirnov- and Ballooning coils) of ASDEX Upgrade \cite{Schittenhelm_1997, Mink_2016, Horvath_2016} are well suited to measure the frequency evolution of the post-disruption modes. The pick-up coils have a sampling frequency of 2~MHz, setting the maximum measurable mode frequency at 1~MHz. While the data acquisition system allows for setting the sampling frequency to 10~MHz for short time intervals, this was not routinely applied.

\section{Mode tracing \label{sec:MT}}

    \begin{figure}[h]
    	\centering
    	\includegraphics[width=\linewidth]{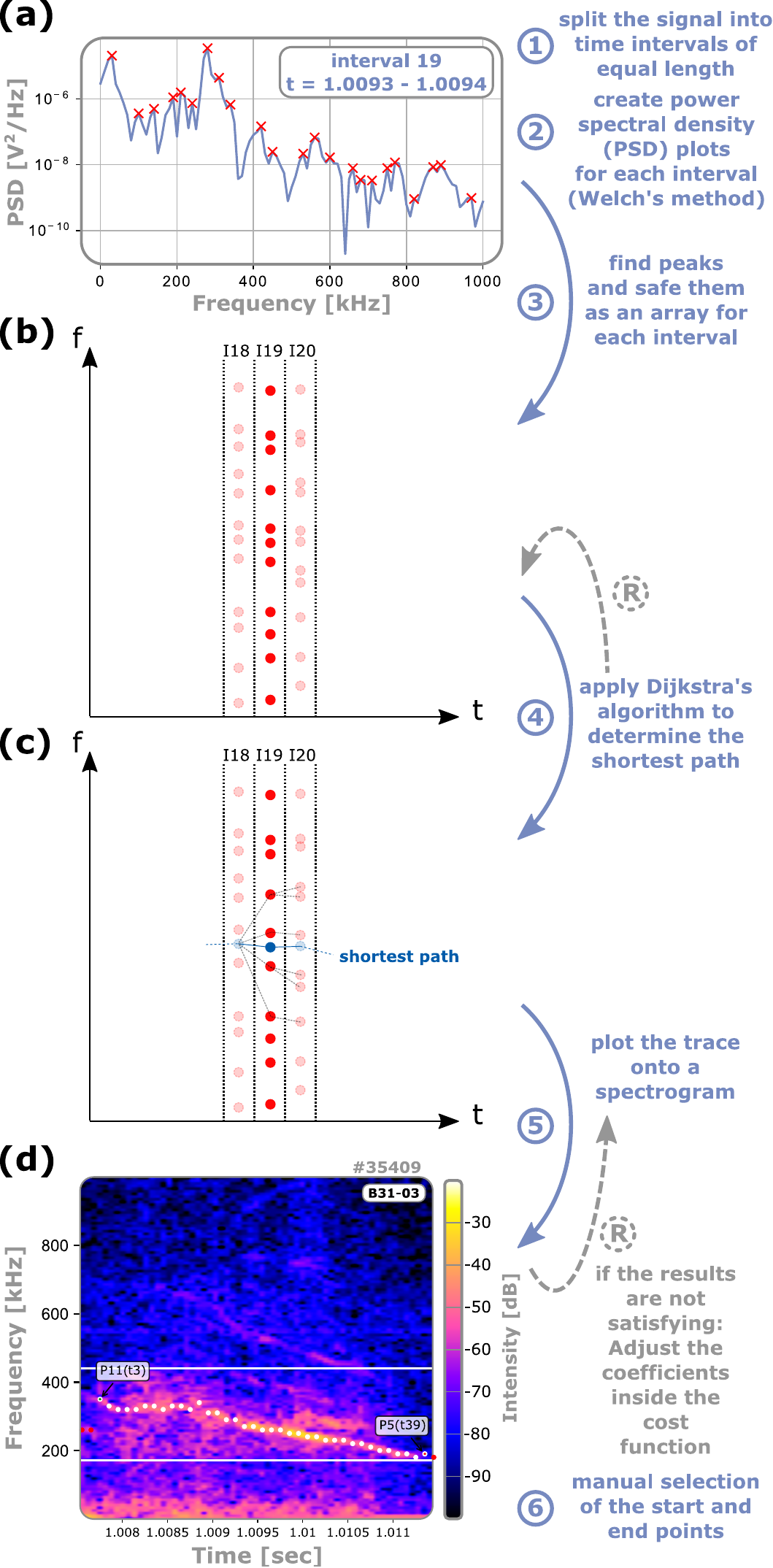}
    	\caption{Concept of the mode tracing algorithm.}
    	\label{fig:MT_algorithm}
    \end{figure}
    
In order to automatically characterise the amplitude and frequency evolution of the modes observed during the current quench, we developed a mode tracing algorithm~\cite{Heinrich_2021_MSc}. The main idea of this algorithm is shown in figure~\ref{fig:MT_algorithm} and briefly summarised in the following.
    
\begin{enumerate}[label=\protect\circled{\arabic*}]
        \item The signal is divided into time intervals of equal length. We used 40 time intervals, which results in a sampling into 40 time-frequency points. This ``downsampling'' is used to reduce computational time.
        \item For each interval, the power spectral density~(PSD) is calculated via Welch's method shown in figure~\ref{fig:MT_algorithm}(a). 
        \item A local peak finding algorithm is applied to create an array of frequencies representing the time evolution of the mode frequency.
        \item The grid necessary to apply Dijkstra's algorithm is initialised. Thereby, each frequency point~(node) from interval $i$ is connected to each other node from interval $i+1$ if the change in frequency \(\Delta f < \Delta f_\textnormal{max}\) is less than a predefined maximum jump frequency. For this values between 0.4 and 6~kHz $\cdot$ 1s/(\(\Delta t_{i \rightarrow i+1}\) in s) were used where 1~kHz was the default setting used in 75 cases.
        The first and the last interval are connected to only a single ``artificial'' node, which act as the start and end points, respectively. To every connection between nodes~(vertex) a value is assigned via the cost function
        
        \begin{equation}
            \textnormal{cost}(i \rightarrow i+1) = c_f \cdot \Delta f + \frac{c_A}{A_{i+1}},
        \end{equation}
        
        with user defined weighting coefficients \(c_f\) and \(c_A\) as well as the absolute amplitude \(A_{i+1}\) measured for the second node in the node-vertex-node connection as suggested by Magyarkuti~\cite{Magyarkuti_TDK_2009}. Ratios for \(c_f\):\(c_A\) between 10:125 and 500:10 were applied, however, a 1:1 ratio was used in 104 cases. Now Dijkstra's algorithm can be applied on this map, which yields the path with the global lowest cost from the start to the end node with one node per interval.
        \item The result is plotted on top of a high-resolution spectrogram (white points in~\ref{fig:MT_algorithm}(d)). If the result is not satisfactory, new values for \(c_f\), \(c_A\) and \(\Delta f_\textnormal{max}\) may be selected \circledtext{R} and the steps \circledtext{4} -- \circledtext{5} are repeated.
        \item When the results are satisfactory, the user can select the start and end points manually in the graphical user interface.
\end{enumerate}

\section{General mode characteristics \label{sec:observations}} 

In this section we are going to discuss general characteristics of the observed modes, such as time-frequency evolution, mode structure, and the time interval in which these appear.

\begin{figure}[h]
        \centering
        \hypertarget{fig1.4}{}
        \includegraphics[width = \linewidth]{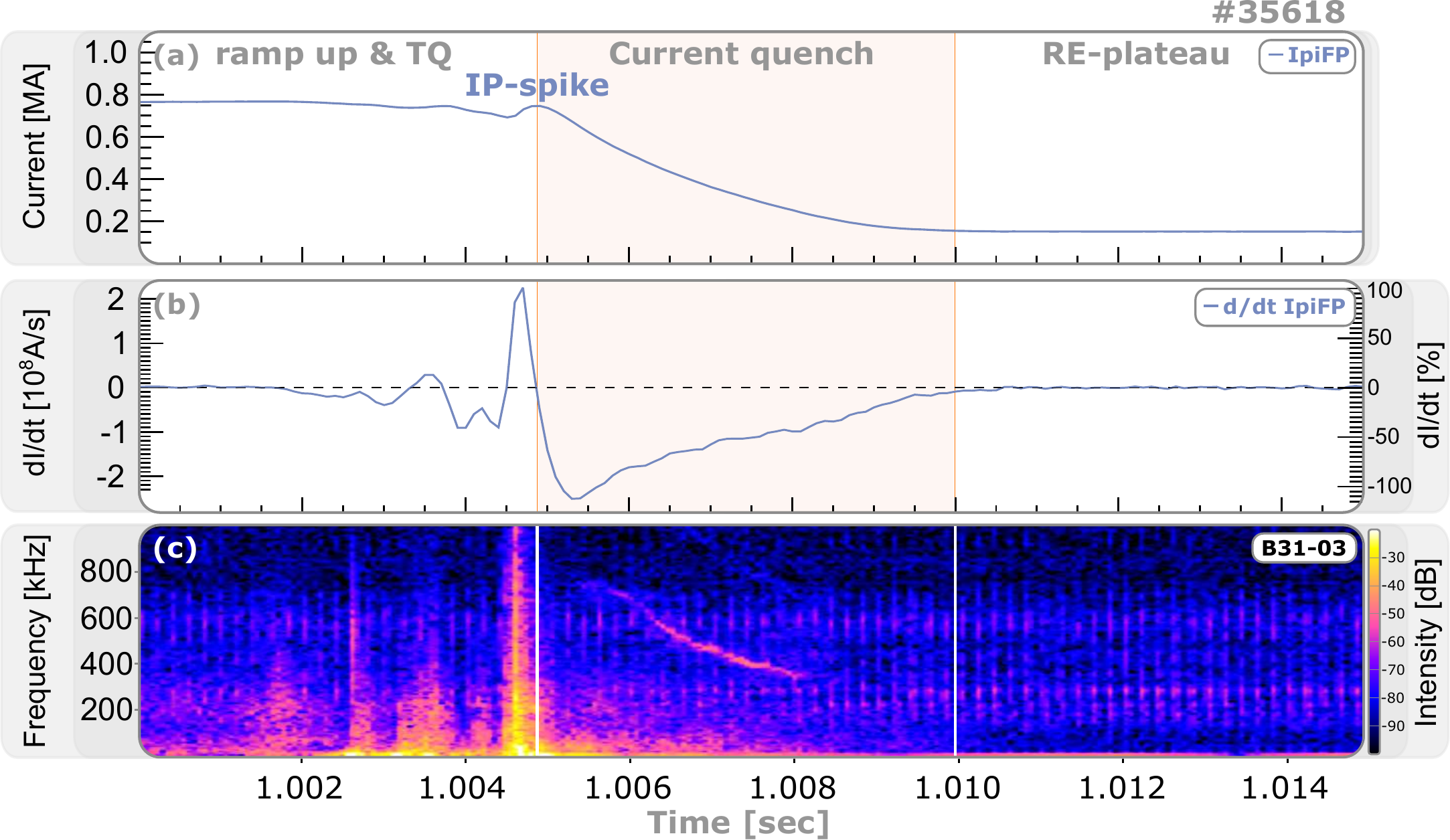}
        \caption{Definition of the CQ time window for the scope of this work. In (a) the evolution of the plasma current is given, with its temporal derivative depicted in (b). In (c) a spectrogram for the magnetic coil B31-03 is provided, which shows the presence of a sweeping mode.}
        \label{fig:definition_CQ}
    \end{figure}

As illustration, the plasma current~\habb{(\(I_\textnormal{p}\))} evolution around the quench phase for discharge \#35618 is displayed in figure~\ref{fig:definition_CQ}(a). The MGI was triggered at 1.000 seconds, and within a few milliseconds the thermal quench~(TQ) occurs. After the $\textnormal{I}_\textnormal{p}$ spike~\cite{Hender_2007_ITER_Physics} the plasma current collapses during the current quench~(CQ) phase, followed by the RE plateau with initial runaway current around 156~kA.
    
    The d$I_\textnormal{p}$/dt temporal derivative of the plasma current is shown in figure~\ref{fig:definition_CQ}(b). For our definition, the initial $\textnormal{I}_\textnormal{p}$ spike marks the start of the CQ, located at the zero-crossing of the temporal derivative. It lasts until the change in plasma current approaches zero again. For the scale on the right in  figure~\ref{fig:definition_CQ}(b), the temporal derivative of the current is normalised to its maximum value. As the consecutive RE plateaus also show a slight decay over time, a threshold at -0.25\% normalised d\(I_\textnormal{p}\)/d$t$ was applied to all discharges to determine a quantitatively reproducible end point of the current quenches.
    
    In figure~\ref{fig:definition_CQ}(c) the spectrogram for the radial magnetic field perturbation coil \mbox{B31-03} is shown. It is located at the low field side midplane and was used as a reference coil for most of the discharges. The spectrogram shows the presence of a mode sweeping from around 800 to 400~kHz between 1.006 to 1.008 seconds, during the current quench.
    
\begin{figure}[h]
    	\centering
    	\includegraphics[width=\linewidth]{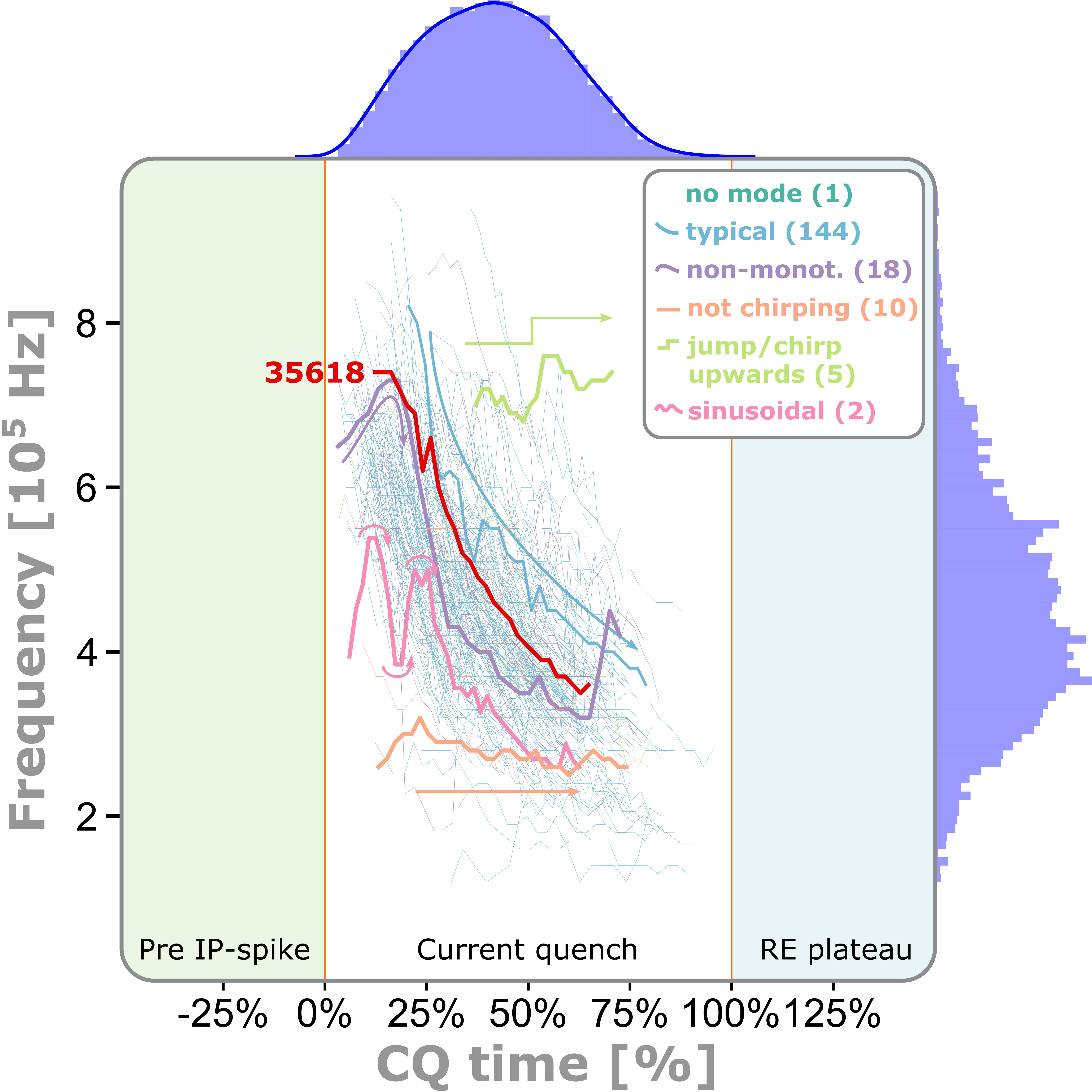}
    	\caption{Time window of CQ mode evolution. The mode frequency traces are plotted as function of time normalised to current quench duration. The modes are located exclusively inside the CQ. The temporal evolution for the reference discharge \#35618 as well as examples for each category are highlighted.}
    	\label{fig:CQ_location}
    \end{figure}

Figure~\ref{fig:CQ_location} is a summary plot of mode evolution in all investigated discharges. The figure shows the frequency evolution -- obtained by the mode tracing algorithm discussed in section~\ref{sec:MT} -- of all modes against time normalised to the current quench time window.
Colours indicate the different sweeping behaviours which will be detailed in section~\ref{sec:Classification}. Discharge \#35618 is highlighted as a reference.
    
While the individual mode evolutions can differ between the discharges, they nevertheless display similar general behaviour.
As shown in figure~\ref{fig:CQ_location}, the time traces of the modes are located exclusively during the current quench phase. For this reason, the modes will be referred to as \highlight{current quench modes}~(CQ mode) within this paper. 
    
    \begin{figure}[h]
        \centering
        \includegraphics[width = \linewidth]{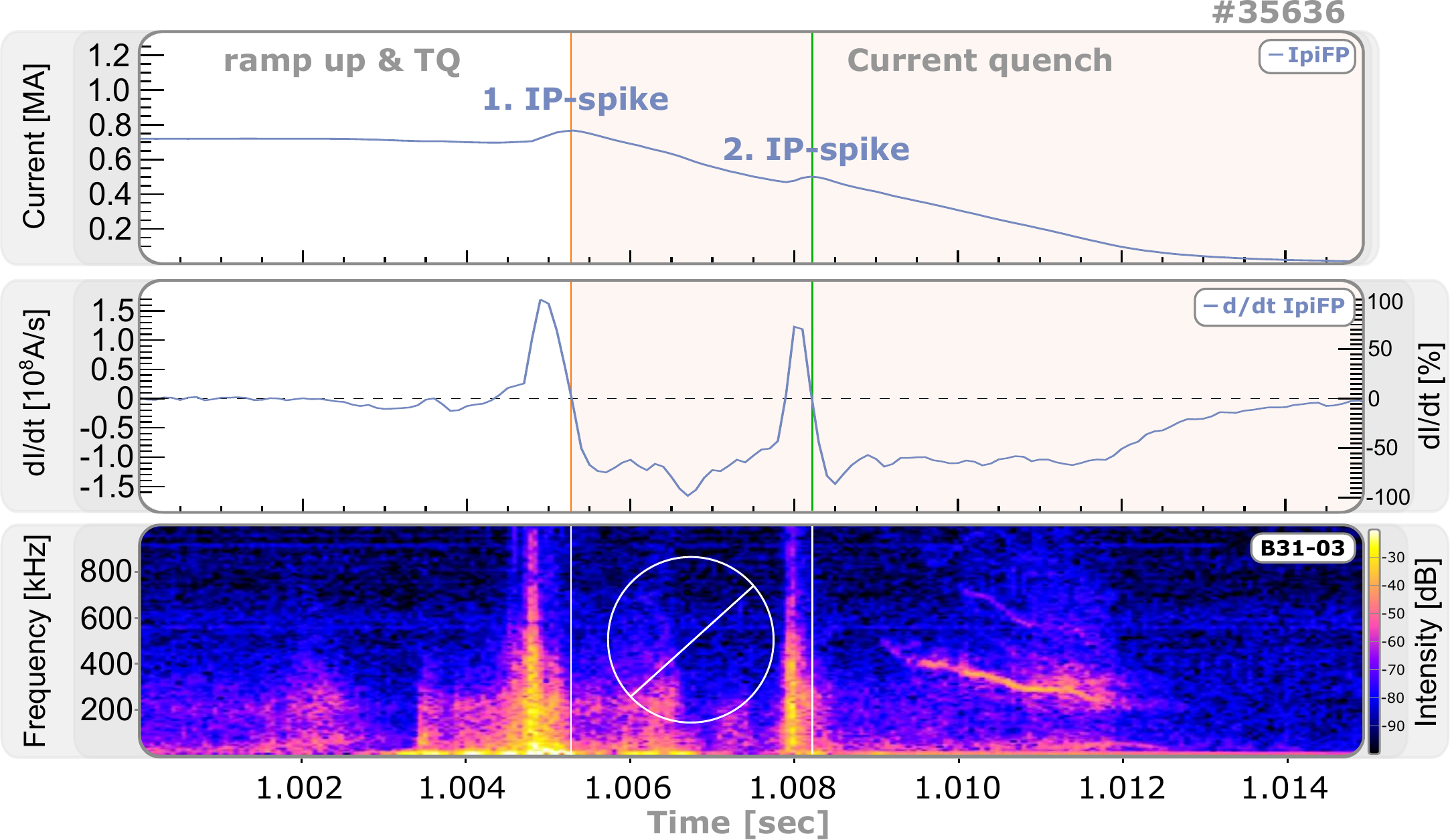}
        \caption{Behaviour of the modes in multi-stage disruptions. The mode always appears after the \textbf{last} $\textnormal{I}_\textnormal{p}$ spike has occurred.}
        \label{fig:incomplete_disruption_35636}
    \end{figure}
    
    Our database includes five discharges, which show incomplete disruptions, i.e. multiple $\textnormal{I}_\textnormal{p}$ spikes appear. In figure~\ref{fig:incomplete_disruption_35636}(a) we observe a second $\textnormal{I}_\textnormal{p}$ spike about 3~ms after the first. As shown in the spectrogram, the mode is only present after the last $\textnormal{I}_\textnormal{p}$ spike has occurred. A similar trend is observed for the other ``multi-stage" disruptions in the database. This indicates that the modes are not triggered by a drop of plasma current alone.
    
    \begin{figure}[h]
        \centering
        \includegraphics[width = \linewidth]{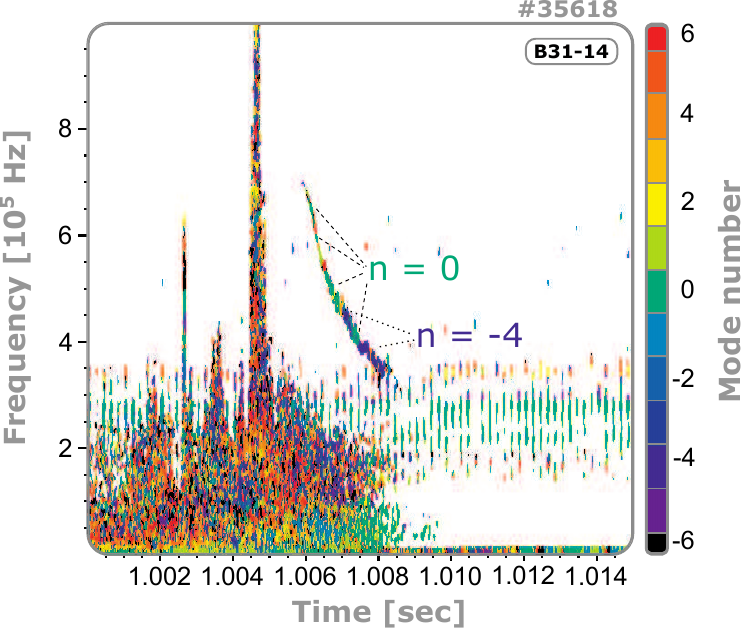}
        \caption{Analysis of the toroidal mode number~\textit{n}. The mode appears to be a mixture of many different mode numbers. The most prominent are \(n = 0\) and \(n = -4\).}
        \label{fig:Mode_number}
    \end{figure}
    
First investigations of the mode location -- details provided in the thesis of P. Heinrich~\cite{Heinrich_2021_MSc} -- point towards strong mode amplitudes in the Soft X-ray channels closer to the separatrix. However, proper normalisation of the signals is highly complex. Therefore, this can only be interpreted as a first hint of the radial location of these modes.
As ASDEX Upgrade is equipped with a large set of magnetic pick-up coils, distributed over the torus in poloidal and toroidal directions, one can calculate the poloidal~(\textit{m}) and toroidal~(\textit{n}) mode numbers \cite{Horvath_2015, Horvath_2016}. Figure~\ref{fig:Mode_number} shows the toroidal mode number calculation for our reference discharge \#35618. Even though the mode is globally coherent, it does not show a unique mode number (neither toroidal nor poloidal) for any of the investigated discharges, but rather a mix of many different ones as shown in the example. Based on the plots, the most prominent toroidal mode numbers are \(n = 0\) and \(n = -4\). A negative \(n\) value represents the propagation of the mode in the ion diamagnetic drift direction.

\section{Classification of the CQ modes \label{sec:Classification}} 
During the investigations of the CQ modes, each of the 179 modes has been characterised and five different categories are identified. In the following a short description of the five characteristics is given.
    
\subsection{Frequency sweeping}
    
\begin{figure}[h]
    	\centering
    	\includegraphics[width=\linewidth]{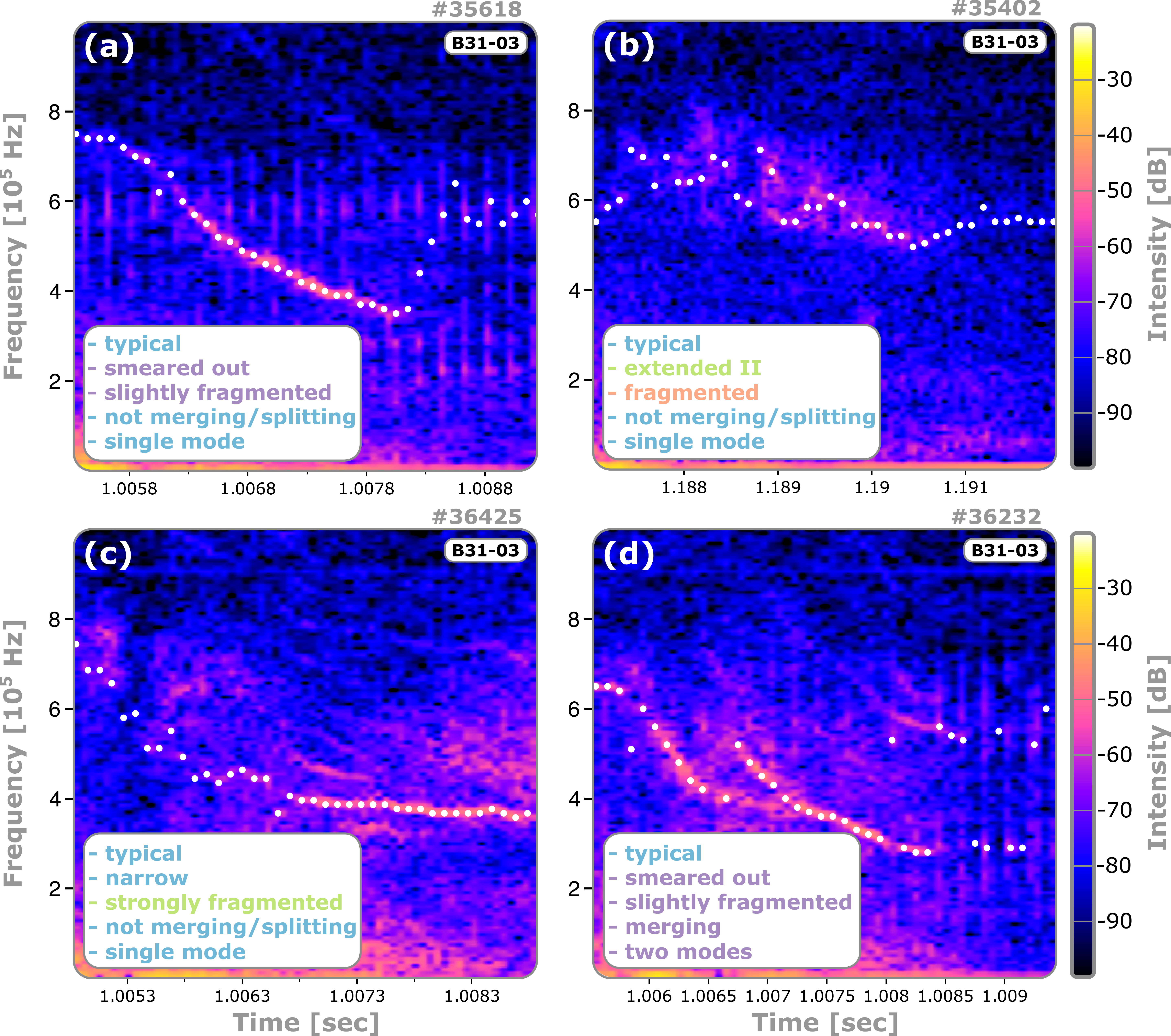}
    	\caption{Comparison of different mode characteristics for the magnetic pick-up coil \mbox{B31-03}.
	The mode characteristics from figure~\ref{fig:Distributions} are given in the boxes for each example.}
    	\label{fig:Spectrograms}
    \end{figure}

The most prominent characteristic of these modes is their frequency sweeping behaviour. As a mode evolves in time, its amplitude and frequency changes. This typically results in a frequency evolution, where the mode starts at higher frequencies -- typically around 800 kHz -- and sweeps towards lower frequencies (see figure~\ref{fig:Spectrograms}(a)). The change in frequency is typically strongest at the beginning and reduces towards the end of the mode. This evolution is the most common type of sweeping behaviour as shown in the distribution in figure~\ref{fig:distribution sweeping}, thus is referred to as~\highlight{typical} throughout this paper. 
Discharge \#35618, representing this kind of sweeping behaviour, is used as a reference and was analysed in more detail, also with the help of simulations (see section~\ref{sec:theory}).
    
    \subsection{Mode contour}
    
    The spectral width of a mode is qualitatively classified as the sharpness of the mode contour. Cases exist -- referred to as~\highlight{narrow} -- which depict a sharp contour, thus have a well defined frequency. The more the modes lose their sharp contour as the spectral width broadens, the harder it is to determine their main frequency. If the main frequency is hard to determine, as the mode does not show a distinct contour, the mode is referred to as~\highlight{extended} (compare figure~\ref{fig:Spectrograms}(b)). If at least a short fragment of the mode shows this behaviour but the rest is well traceable, it is labeled as extended I. In case the mode contour does not show the distinct contour but shows a trend that is visible (compare figure~\ref{fig:Spectrograms}(b)) it is referred as extended II and in case this trend is not visible anymore as extended III.
    \subsection{Degree of fragmentation}
    
    Modes can also show fragmentation effects which are classified by the~\highlight{degree of fragmentation}. It indicates, how continuously a mode evolves in time and how many intersections appear. The higher the degree of fragmentation, the harder it is to trace the amplitude and frequency evolution of the modes. This parameter also depends on the position of the magnetic coil. The distribution given in figure~\ref{fig:distribution fragmentation} shows the degree of fragmentation mainly for the reference coil B31-03. In a few cases better tracking precision can be achieved by using other pick-up coils or soft X-ray.
    
    \subsection{Merging \& splitting of modes}
    In some discharges modes are present, which seem to split into two mode branches or emerge from them. Separate mode branches are only counted as a second mode if they are clearly separated. If two modes appear within a discharge, they mostly merge at a certain point in time (see figure~\ref{fig:Spectrograms}(d)). For most discharges we observe no merging or splitting behaviour as shown in figure~\ref{fig:distribution merging}.
    
    \subsection{Number of modes}
    
\begin{figure}[h]
    	\centering
    		\subfigure[Distribution of different sweeping behaviours.]{\label{fig:distribution sweeping}\includegraphics[width=\linewidth]{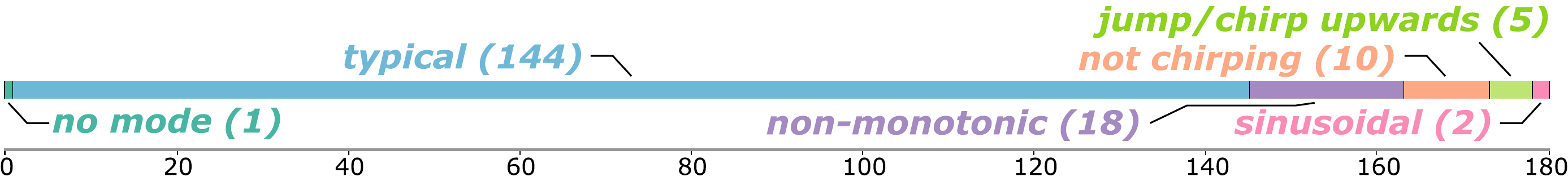}}
    		\subfigure[Distribution of different mode contours.]{\label{fig:distribution contour}\includegraphics[width=\linewidth]{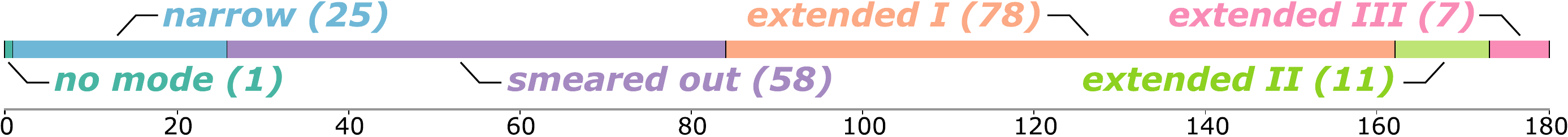}}
    		\subfigure[Distribution of different degrees of fragmentation.]{\label{fig:distribution fragmentation}\includegraphics[width=\linewidth]{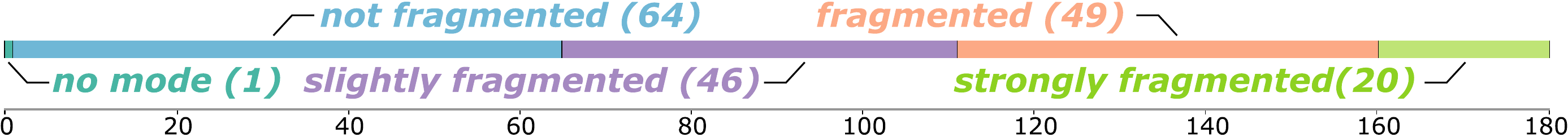}}
    		\subfigure[Distribution of merging and splitting of modes.]{\label{fig:distribution merging}\includegraphics[width=\linewidth]{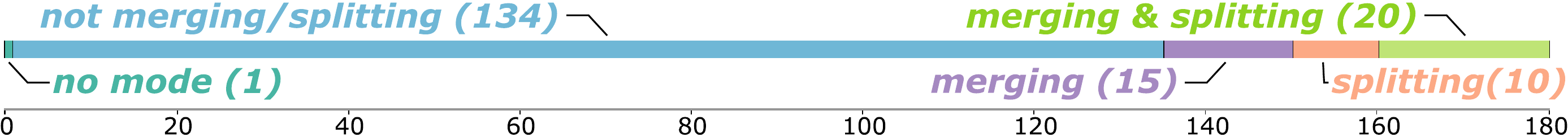}}
    		\subfigure[Distribution of number of modes which appear inside each discharge.]{\label{fig:distribution number modes}\includegraphics[width=\linewidth]{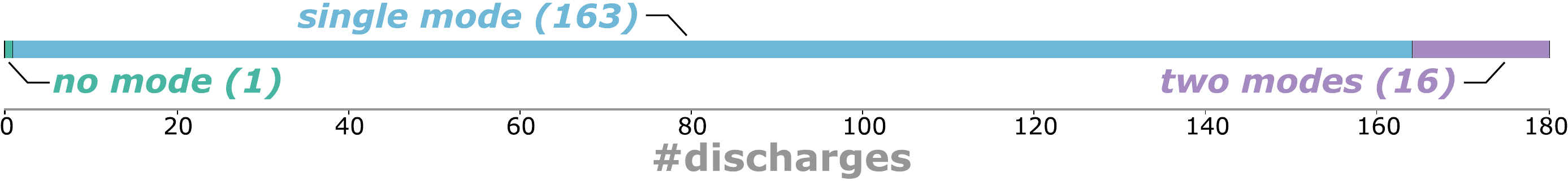}}
    	\caption{Distributions of the five mode characteristics with the total number inside the parentheses.} 
    	\label{fig:Distributions}
    \end{figure}

    As shown in figure~\ref{fig:distribution number modes}, in most discharges we observe single modes and often their upper harmonics. For this classification the number of harmonics is not taken into consideration, thus only the number of distinct modes was investigated. For 16 discharges in our database we observe two modes. Nearly all of them merge at a certain point in time -- between approximately 30 -- 55\% of the total duration of merging modes.
    The only discharge, which seems not to contain any modes is \#38074. Neither in the magnetics nor in the soft X-ray lines of sight~(LOSs) any activity was observed. However, the global parameters do not show any prominent differences to the other discharges from the same session.
    
\section{Statistical analysis \label{sec:statistical_analysis}} 
With the help of the mode time traces and characteristics, multiple statistical analysis methods had been applied and hypotheses tested. It was investigated whether different mode parameters (e.g. the above mentioned characteristics, the mode amplitude or duration) show an effect on the subsequent RE formation. In subsection~\ref{subsec:MLFs}, the results of a histogram based analysis to determine the modes' most likely frequencies is given. A correlation matrix for 38 different observables was analysed in detail and the results are summarised in subsection~\ref{subsec:CM}. In subsection~\ref{subsec:FA}, the results from the factor analysis are presented, which aims to connect the observables of a system with underlying (hidden) factors. The results from hypothesis testing with the help of regressions are given in subsection~\ref{subsec:Regressions}.

    \subsection{Most likely frequency \label{subsec:MLFs}}
\begin{figure}[h]
    	\centering
    	\includegraphics[width=\linewidth]{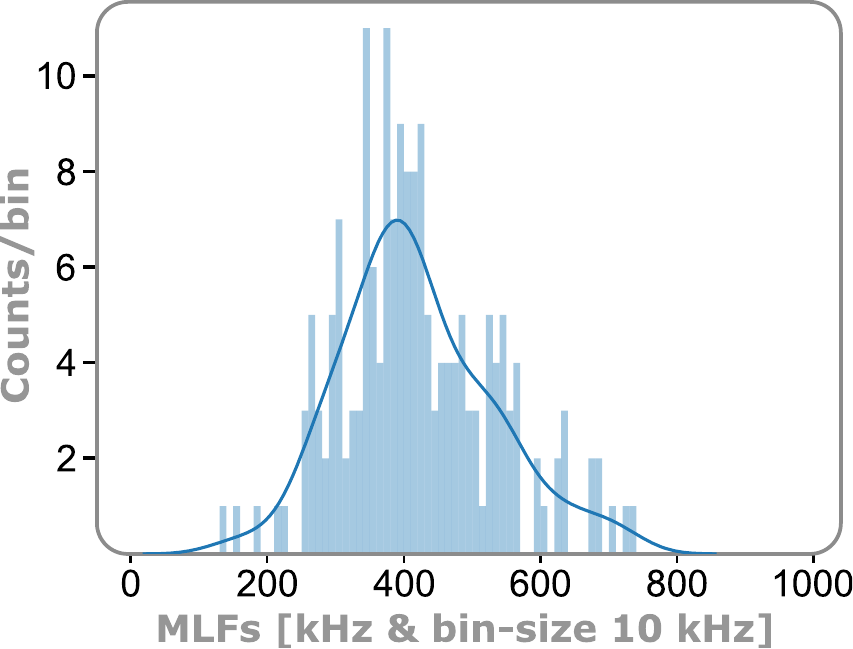}
    	\caption{The distribution of the most likely frequencies (MLFs) of the modes is shown with maxima at 340 and 370~kHz.}
    	\label{fig:MLFs}
    \end{figure}

    In case the CQ modes have a significant effect on the formation of RE plateaus, a future external mode drive might become desirable. For the excitation of these waves, their resonance frequency has to be known in advance. Based on our database the so-called \highlight{most likely frequency}~(MLF) for each discharge was determined. It represents the mode frequency, which is maintained for the longest duration. Introduced by the mode tracing algorithm, a time sampling of 40 time points is applied and the frequencies within 10~kHz-bins are counted. As these modes often sweep strongest at the start, the MLFs are not necessarily given by the arithmetic mean. Therefore, the individual MLF for each discharge was determined via histograms with a bin size of 10~kHz. For every discharge, the individual probability density function of the mode frequency was calculated. The MLF is determined by the global maximum of this distribution, thus the most probable frequency. The distribution of the 179 MLFs for each mode is shown in figure~\ref{fig:MLFs}. The maxima are located at 340 and 370 kHz. Therefore, the most probable resonance frequency for externally driven modes is slightly below~400~kHz.
    
    \subsection{Summary of the correlation matrix analysis \label{subsec:CM}}
    A correlation matrix for 38 observables was calculated and analyzed in detail. This includes for example the plasma and subsequent RE current, toroidal magnetic field strength, plasma temperature, the plasma safety factor at \(\rho_\textnormal{pol}= 0.95\)~(q95) or density measurements.
    A detailed list and description for each observable as well as the correlation matrix itself is given in~\ref{sec:appendix}.1 and 2. In the following the results of the analysis are presented.
    Most of the higher correlations between the plasma parameters can be explained by well known interactions (e.g. the connection of the plasma current with the plasma safety factor).
    In addition to those correlations two regions of special interest can be observed in the matrix. The relevant excerpt of this correlation matrix discussed in the following is provided in table~\ref{tab:CM_excerpt}.
    
    \begin{table*}
	\caption{Excerpt of regions of interest in the correlation matrix given in~\ref{sec:appendix}.1 and 2. The description for all quantities are available in the table in~\ref{sec:appendix}.1. In bold, correlation values with an absolute value above 0.3 are presented. No strong correlation between mode characteristics defined in section~\ref{sec:Classification} and global plasma/disruption parameters are observed.}
        \label{tab:CM_excerpt}
        \centering
        \def\arraystretch{1.5} 
        \begin{tikzpicture}
            \node (table) [inner sep=0pt] {
            \begin{tabular}{c|clclclclclclc}
                \diaghead{xxxxxxxxxxxxx}{\hspace{.3cm}mode}{plasma} & $\textnormal{N}_\textnormal{Z} $ [bar] & dI [kA] & $\textnormal{I}_\textnormal{RE}$ [kA] & $\textnormal{dt}_\textnormal{CQ}$ [s] & $\textnormal{q}_\textnormal{95}$ & $\textnormal{I}_\textnormal{P0}$ [kA] & $\textnormal{W}_\textnormal{E}$ [kJ] & $\textnormal{T}_\textnormal{e}$ [keV]\\
                \hline
                sweeping & 0.0 & -0.03 & +0.05 & -0.04 & -0.02 & +0.02 & +0.19 & -0.02\\
                \hline
                contour & +0.04 & +0.09 & -0.08 & +0.05 & 0.00 & +0.06 & -0.06 & -0.13\\
                \hline
                fragment. & +0.03 & -0.07 & +0.11 & -0.08 & 0.00 & +0.03 & +0.11 & -0.12\\
                \hline
                merg./split. & +0.01 & -0.13 & +0.07 & -0.11 & +0.04 & -0.12 & -0.07 & -0.07\\
		\hline
                \#modes & -0.02 & -0.14 & +0.15 & -0.16 & +0.02 & -0.03 & -0.05 & 0.01\\
                \hline
                $\textnormal{dt}_\textnormal{mode}$ [s] & \textbf{-0.43} & \textbf{+0.44} & \textbf{-0.46} & \textbf{+0.50} & -0.07 & +0.17 & -0.15 & -0.03\\
                \hline
                $\textnormal{f}_\textnormal{MLF}$ [kHz] & -0.08 & +0.20 & +0.01 & +0.12 & -0.15 & \textbf{+0.34} & +0.26 & +0.03\\
                
            \end{tabular}
            };
            \draw [rounded corners=.5em] (table.north west) rectangle (table.south east);
        \end{tikzpicture}
        
    \end{table*}

    \subsubsection{Correlation between CQ- and mode-times with global plasma parameters (region 2)}\label{subsec:CorrMat2}
    In this first region of interest, the timing and length of the CQ and CQ modes are connected to global plasma parameters such as the (runaway) current, toroidal magnetic field strength, q95 and MGI gas quantity (compare row $\textnormal{dt}_\textnormal{mode}$ in table~\ref{tab:CM_excerpt}).
    The modes appear exclusively during the CQ as shown in section~\ref{sec:observations}. For longer current quench and mode duration, a lower subsequent current is held by runaway electrons~(\(I_\textnormal{RE}\)) and a higher change in current~(d\(I\)) will be achieved as the current change rates~(d\(I_\textnormal{p}\)/dt) are very similar between each shot. This results in high correlation values -- positive (correlated) for the change in current, negative (anti-correlated) for the runaway current. Another observation of this region is the anti-correlation between the amount of gas injected to trigger the disruption~(\(\textnormal{N}_\textnormal{Z}\)) with the CQ and mode times ($\textnormal{dt}_\textnormal{CQ}$ and $\textnormal{dt}_\textnormal{mode}$) as well as the start~(\(t_\textnormal{IP} - t_\textnormal{trigger}\)) and end~(\(t_\textnormal{end CQ} - t_\textnormal{trigger}\)) times of the current quench. More gas causes hereby shorter CQ time windows in which the mode can exist. The values for the safety factor~\(q_{95}\) are typically adjusted by varying the plasma current and toroidal magnetic field. As the pre-disruption plasma current is determined by the timing of the MGI, the trigger time shows strong correlations with the toroidal magnetic field component~(\(B_\textnormal{tor}\)) and \(q_{95}\). Thereby, earlier injections lead to lower initial plasma currents, hence higher safety factors by design.
    
    \subsubsection{Mode characteristics and global plasma parameters (region 4)}\label{subsec:CorrMat4}
    The region in the correlation matrix which connects the mode characteristics from section~\ref{sec:Classification} with the global plasma parameters was the main focus of this investigation (excerpt given in table~\ref{tab:CM_excerpt}). A high correlation would point towards a connection between the mode behaviour and global parameters. However, we do not observe any significant correlations in this region. The subtle correlations within this regions can be explained due to the nature of the characterisation system used within the mode tracing algorithm~\cite{Heinrich_2021_MSc}. Consequently, the mode characteristics and amplitudes do not seem to affect the -- or being affected by -- global plasma parameters.

    \subsection{Results from factor analysis \label{subsec:FA}}
\begin{figure}
    	\centering
    	    \subfigure[Example scree plot showing an inflection point]{\includegraphics[width=0.7\linewidth]{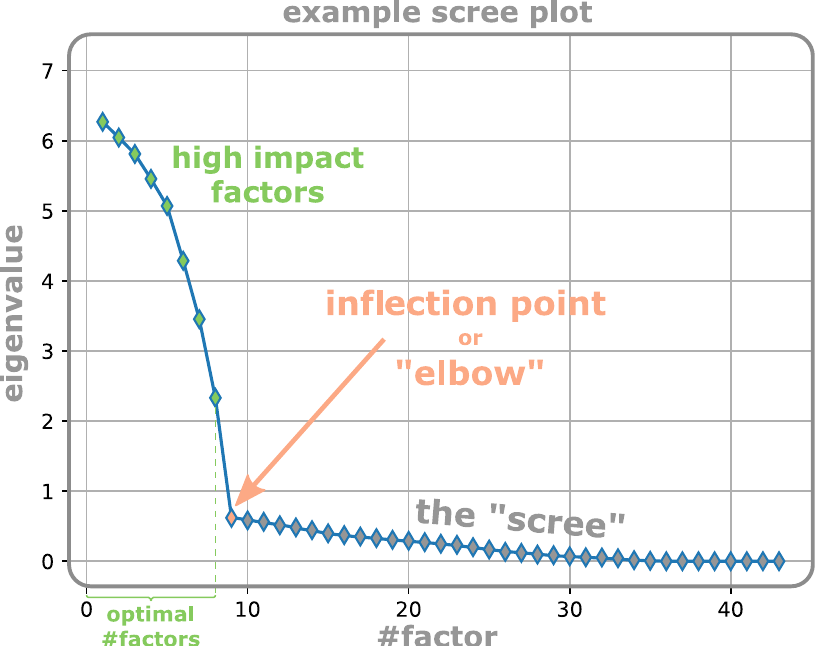}\label{fig:example scree plot}}
    	    \hfill
    	    \subfigure[Scree plot of the FA for the database.]{\includegraphics[width=0.7\linewidth]{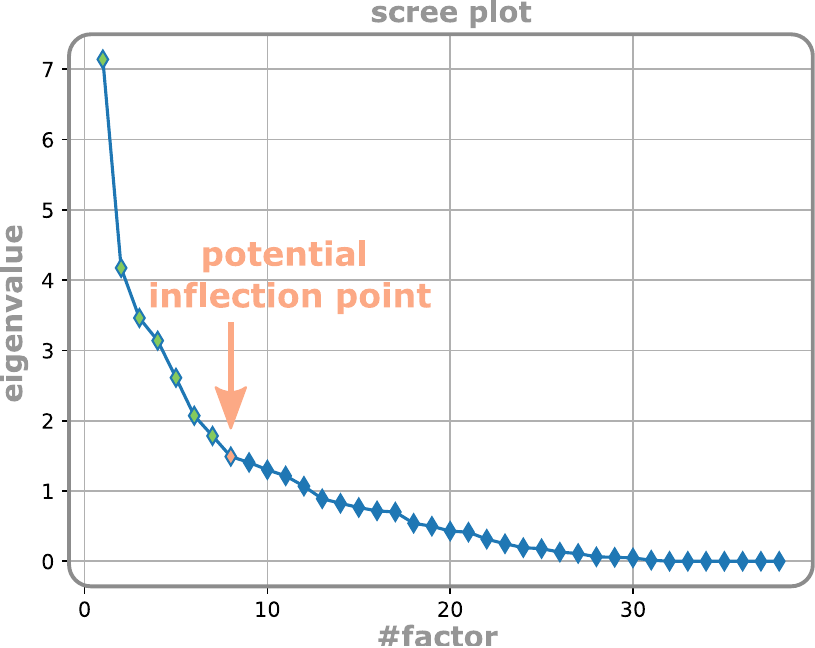}\label{fig:scree plot}}
    	\caption{In~(a) an example for a typical scree plot is given. The inflection point separates the high impact factors from the ''scree''. In~(b) the scree plot of the EFA for our database is depicted. It does not show a clear inflection point, being the usual, strong separation between factors. Factor~8 is highlighted as a potential inflection point.}
    	\label{fig:scree plots}
    \end{figure}

    In this concept multiple observables of the system are reduced to a few underlying factors~\cite{Yong2013}. These factors are also often referred to as hidden variables as they are not directly measured, but their effect can be visible throughout the observables.
    A scree plot is used to determine how many factors are necessary to describe the underlying connections of the system. It visualizes the separation of high impact factors for the system, from the rest (''scree''). It typically shows a so-called elbow or inflection point depicted in figure~\ref{fig:example scree plot}, which acts as the desired separation between factors.
    
\begin{figure}[htb]
        \centering
        \includegraphics[width=\linewidth]{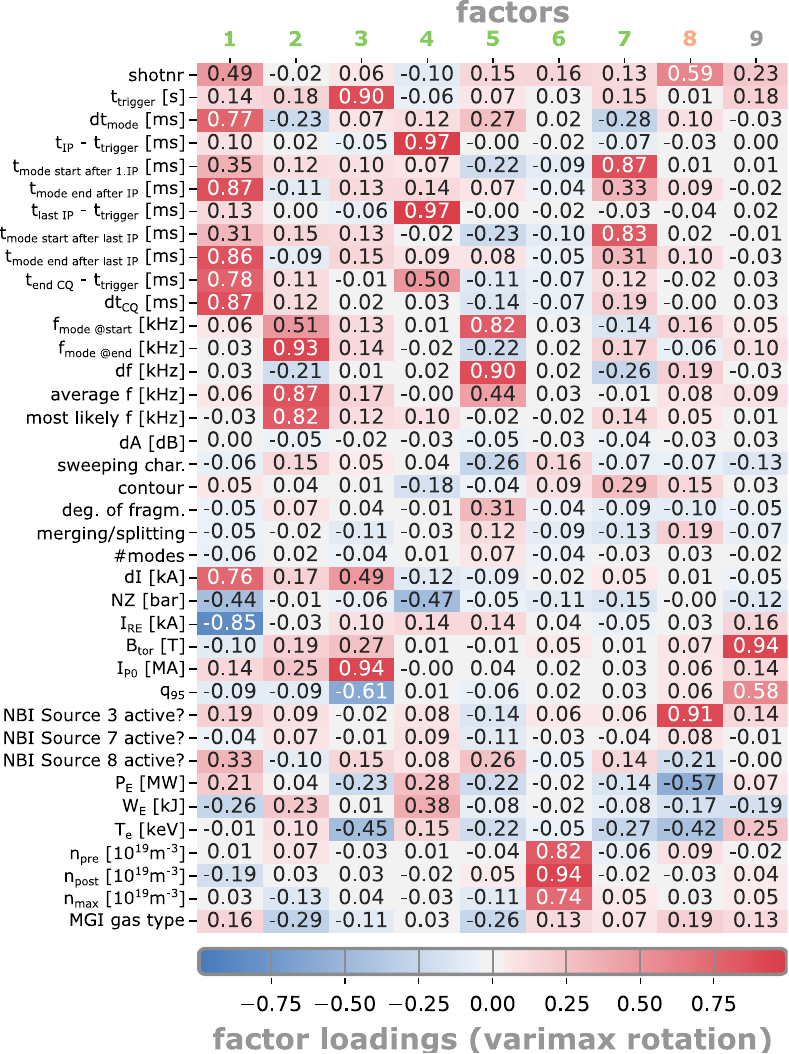}
        \caption{The loadings for the first 9 factors resulting from the factor analysis. A summarizing description for each factor is provided in table~\ref{tab:factor_names}. The orange color of factor 8 is used to mark it as the potential weak inflection point.}
        \label{fig:FA}
    \end{figure}

    In our case we used exploratory factor analysis~(EFA) to find factors, which describe a possible interaction between global parameters and mode characteristics. The results presented in figures~\ref{fig:scree plot} and~\ref{fig:FA} were created with the python module \texttt{FactorAnalyzer}~\cite{FactorAnalyzer} with the rotation-mode \texttt{varimax} (more details in the documentation~\cite{FactorAnalyzer}). The scree plot in figure~\ref{fig:scree plot} does not show a distinct inflection point. This indicates, that the results could not be reduced to a few underlying factors. This could be due to the system being highly complex or the observables not being suited to describe the system well enough with only a few factors. The first 9~factors and their loadings (values indicating the effect of that variable on the factor) are shown in figure~\ref{fig:FA}. Hereby, the factors are sorted by eigenvalues, thus their impact on the system.
    The factor analysis code can not provide a description or name for each factor. Consequently, we will refer to each factor by it's number displayed at the top.
    A detailed analysis and interpretation for the loadings of each factor given in figure~\ref{fig:FA}, is available in the~\ref{sec:appendix}.3.
    
    As a result of the analysis process, custom names for each factor are given in table~\ref{tab:factor_names}. Our main focus was to analyse connections between mode characteristics and other observables via hidden variables of the system. However, also with this analysis technique, no significant connections have been observed. Most of the loadings were already expected due to prior knowledge of the physical dependencies or experimental setup, as indicated in the table~\ref{tab:factor_names}. Again, higher loadings for the mode characteristics can be connected to the tracing and characterisation methods instead of physical phenomena.
    
    \begin{table}[tb]
        \caption{Potential custom names and description of the factors based on their loading.}
        \label{tab:factor_names}
        \centering
        \begin{tikzpicture}
            \node (table) [inner sep=0pt] {
            \begin{tabular}{l|c|p{0.4\linewidth}}
                 & \textbf{Custom name} & \textbf{Factor description} \\
                \hline
                1 & CQ-extent & determines the subsequent RE current\\
                \hline
                2 & frequency-plateau & flattening of the mode frequency sweep towards the end of the mode\\
                \hline
                3 & MGI trigger timing & sets post-disruption plasma parameters, such as plasma current\\
                \hline
                4 & gas penetration & amount of MGI gas injected directly affects the disruption times\\
                \hline
                5 & steep mode chirp & mode sweeping behaviour at the start affects the change in frequency and average frequency\\
                \hline
                6 & density measurements & plasma density measurements\\
                \hline
                7 & mode delay & timing of the mode start with respect to the (last) $\textnormal{I}_\textnormal{p}$ spike\\
                \hline
                8 & experimental setup & routine of the discharges varied over time\\
                \hline
                9 & magnetic setup & connection between \(B_t\) and \(q_{95}\)
            \end{tabular}
            };
            \draw [rounded corners=.5em] (table.north west) rectangle (table.south east);
        \end{tikzpicture}
        
    \end{table}
    
    \subsection{Regressions}\label{subsec:Regressions}
\begin{figure*}[ht]
    	\centering
    	\includegraphics[width=\linewidth]{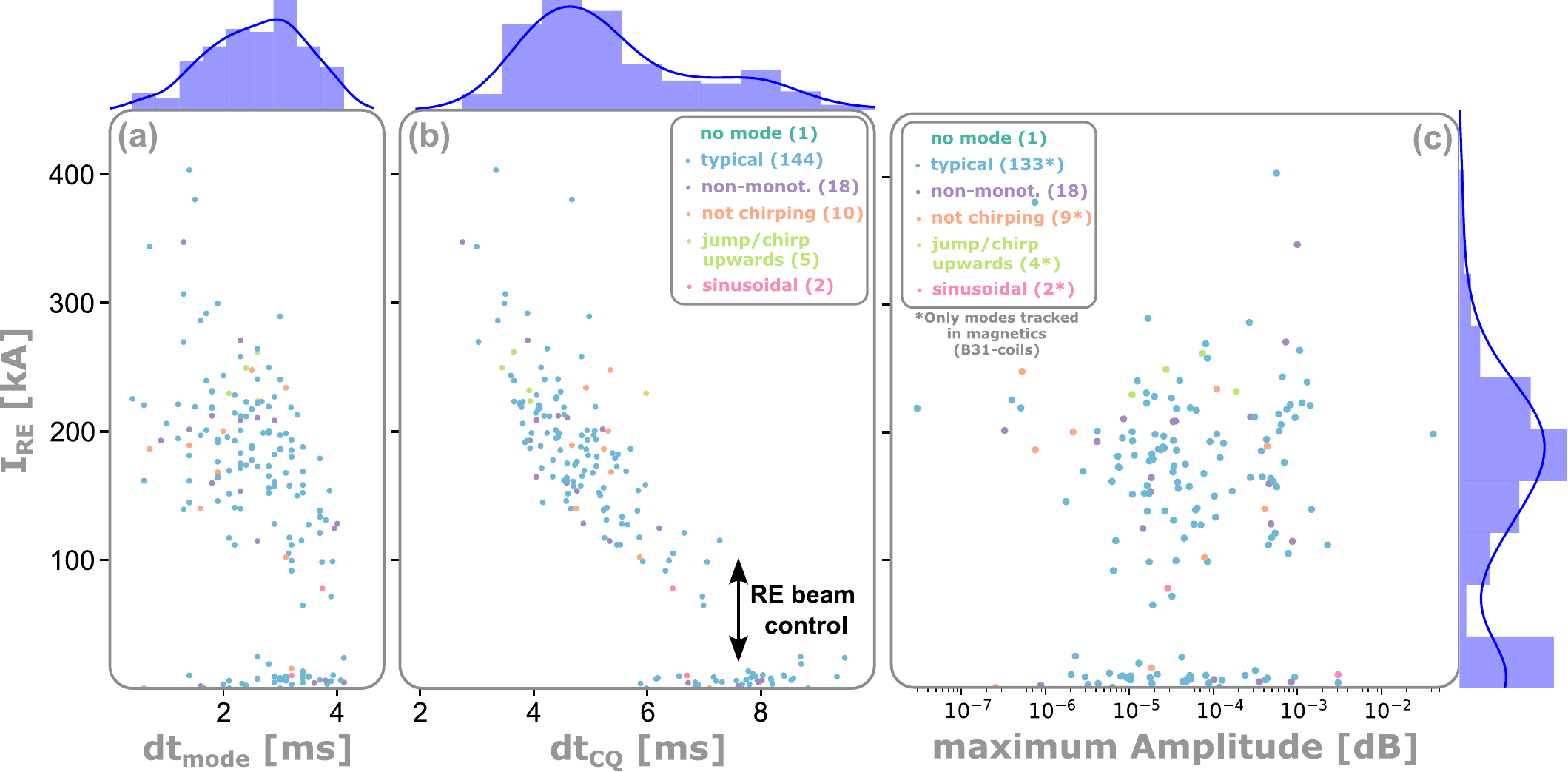}
    	\caption{Runaway current \(I_\textnormal{RE}\) plotted against the mode duration in~(a), current quench duration in~(b) and mode amplitude in~(c). No effect of the mode amplitude on the RE beam current is observed in~(c).}
    	\label{fig:I_RE plots}
    \end{figure*}

    In the contribution by Lvovskiy \textit{et al.}~\cite{Lvovskiy_2018, Lvovskiy_2023}, the duration of MHD activity and the loss of the initial RE seed are connected, thus the consequent runaway current \(I_\textnormal{RE}\) is affected. To analyse this connection for ASDEX Upgrade, linear regressions are created.
    In figure~\ref{fig:I_RE plots}(a) the RE current is shown against the duration of the current quench modes. The trend shows a decreasing runaway current with increasing mode activity time, supporting the hypothesis. Interestingly, we do not observe modes, which last longer than \({\sim}4\ \textrm{ms}\), even though the current quench lasts up to about 10~ms as shown in figure~\ref{fig:I_RE plots}.
    On the other hand, in figure~\ref{fig:I_RE plots}(b) the runaway current is plotted against the duration of the current quench, which seems to set the boundary conditions for the appearance of these modes. (The minimum current required to maintain position control sets the lower limit on RE beam current.) In direct comparison of these two plots, the data in~(b) seems to be less scattered. However, the correlation in figure~\ref{fig:I_RE plots} might be also induced by a different causal link, explained in the following. The longer the duration of the current quench the larger the time window of potential mode activity and also more time for the plasma current to drop to lower values during the current quench. In other words, lower runaway currents are typically preceded by longer current quenches, which leaves more time for the modes to exist. A causal relation in either way is difficult to show just from the correlation.
    
    Consequently, how the CQ modes affect the effective d\(I_\textnormal{p}\)/dt during the CQ might be the more important parameter in this case. However, d\(I_\textnormal{p}\)/dt is very similar for the discharges within the data set.
    In addition, we only observed one discharge, where we could not observe a mode, while the global plasma parameters do not significantly differ from the other discharges.

Most importantly, as shown in  figure~\ref{fig:I_RE plots}(c), we do not observe a decreased runaway current for increased mode amplitudes. The points seem to be more scattered, and there is no clear trend that one could identify between mode amplitude and runaway beam current. This shows that there appears to be little causal relation between the two in either direction. The REs are not expected to undergo resonant wave-particle interaction with MHD modes at the observed frequencies of a few hundred kHz. It also appears that the modes don't have sufficient amplitude to perturb the field enough for transport to overcome RE generation.

\section{Simulation analysis} \label{sec:theory}
    
While compressional Alfv\'en eigenmodes~(CAEs) are used as a possible explanation for the observations at \mbox{DIII-D}~\cite{Lvovskiy_2018, Lvovskiy_2023, Liu_2021, Liu_2023} the modes observed on ASDEX Upgrade appear in different frequency ranges. We therefore propose a different potential driving mechanism for the post-disruption modes, which are identified as \highlight{global Alfv\'en eigenmodes}~(GAEs).
    
\begin{figure}[htb]
        \centering
        \includegraphics[width=\linewidth]{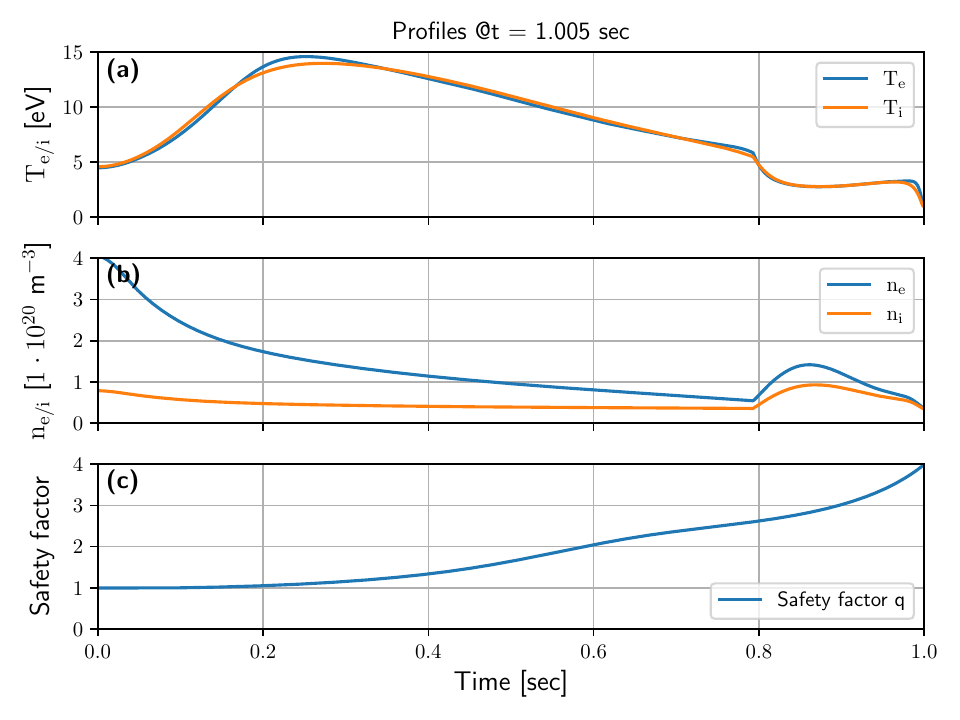}
        \caption{Excerpt of the plasma profiles used as input for the LIGKA simulations. The temperature~(a) and density~(b) profiles for electrons and ions as well as the q-profile~(c) is provided.}
        \label{fig:profiles}
\end{figure}

\begin{figure}[htbp]
    	\centering
    		\subfigure[LIGKA results for \(t = 1.005\ \textrm{s}\). The gray arrows show the potential radial extent of the \(m = 2\) and \(m = 3\) mode.]{\includegraphics[width=0.8\linewidth]{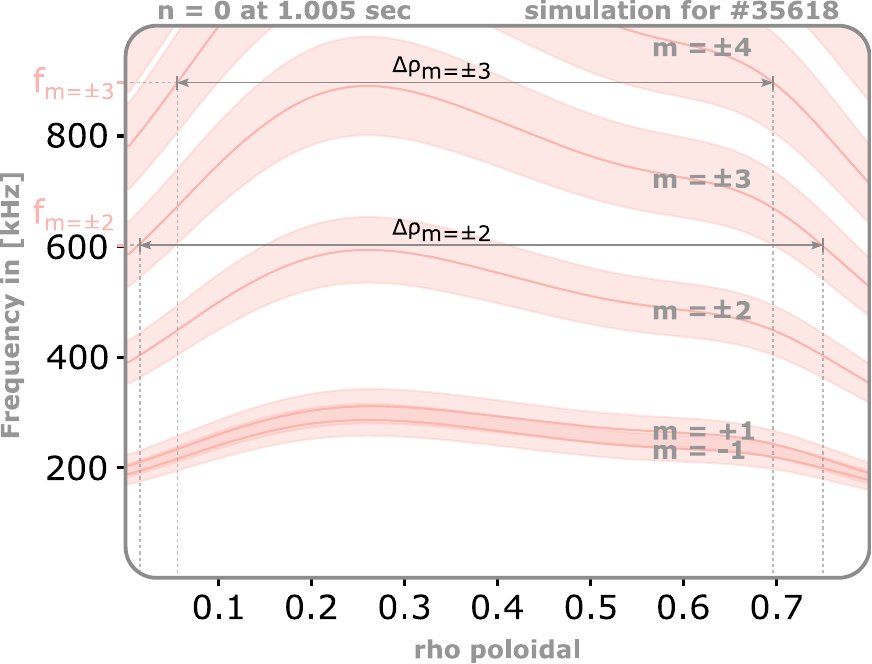}\label{fig:n0_1005}}
    		\subfigure[LIGKA results for \(t = 1.008\ \textrm{s}\). The dashed lines (\(\textnormal{f}_{m = \pm 2/3}^{time}\)) represent the \(m = 2\) and \(m = 3\) maxima for \(t = 1.005, 1.006\ \textnormal{and } 1.007\ \textrm{s}\), respectively. ]{\includegraphics[width=0.8\linewidth]{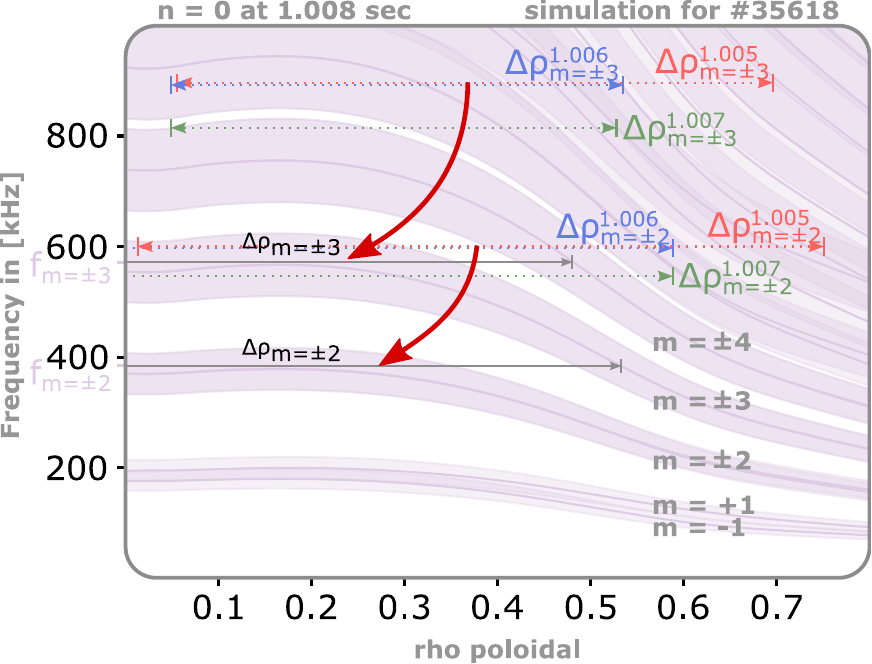}\label{fig:n0_1008}}
    		\subfigure[Spectrogram for magnetic pick-up coil \mbox{B31-03} with the frequency evolution from plots \mbox{(a) -- (e)}.]{\includegraphics[width=0.8\linewidth]{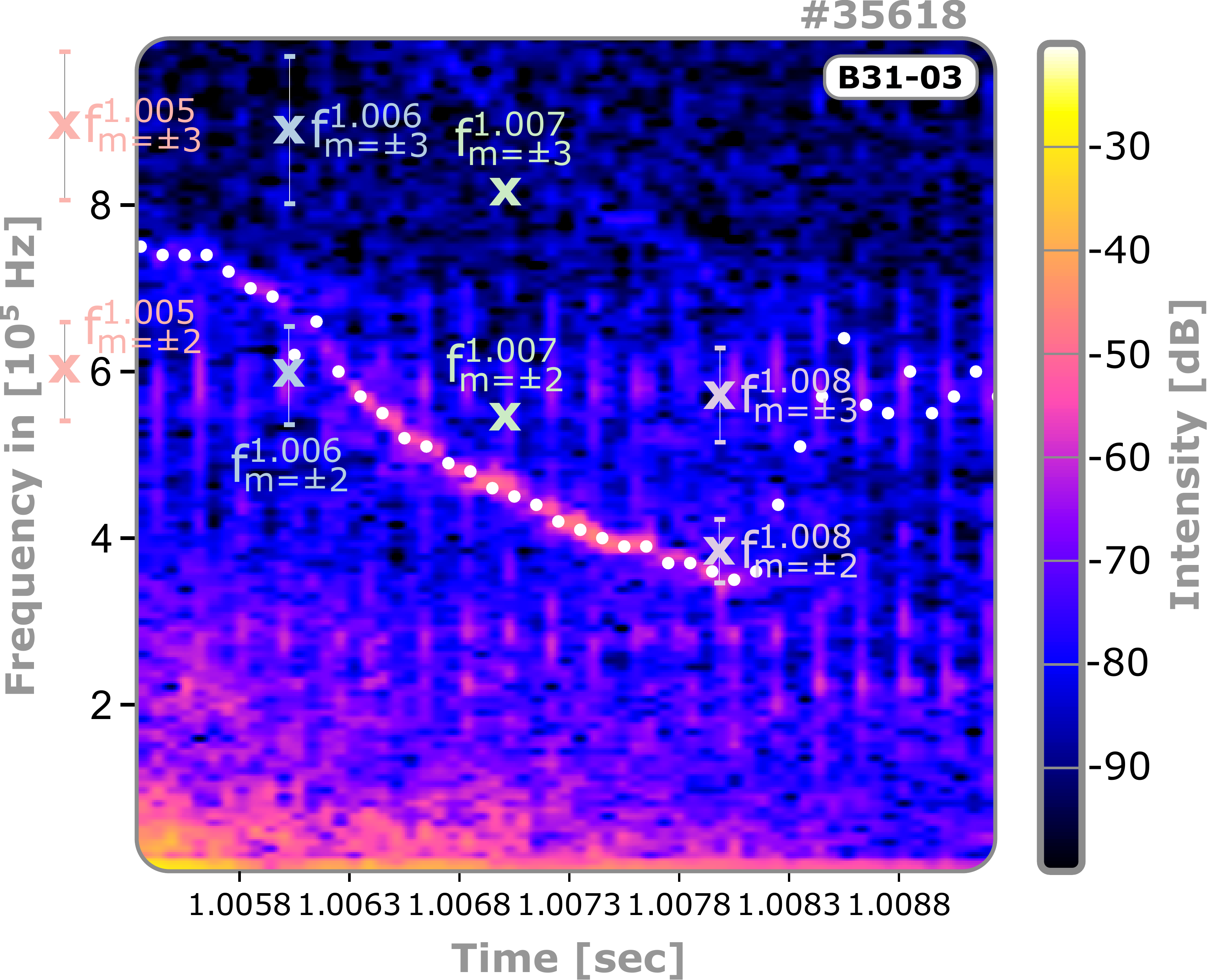}\label{fig:theory_spec}}
    	\caption{Results from LIGKA simulations for discharge \#35618. The plots in~(a) and~(b) show the Alfv\'en continuum. GAEs can be exited above the maxima, thus their frequency~(f) and potential radial extent~(\(\Delta\rho\)) can be estimated. The expected frequency evolutions are compared to the experimental result in~(c).}
    	\label{fig:GAE}
    \end{figure}

We performed ASTRA simulations~\cite{Linder_2020}  for the reference discharge \#35618. These simulations provide local plasma parameters such as the \(q\)- or the electron and ion temperature profiles. Together with the equilibrium reconstruction~\cite{Mc_Carthy_1999, Dunne_2012}, the data was fed into LIGKA~\cite{Lauber_PHD, Lauber_2007} simulations.
In figure~\ref{fig:profiles} the temperature and density profiles for electrons and ions is provided as well as the safety factor~(q) profile.
    LIGKA (LInear GyroKinetic shear Alfv\'en physics) is an eigenmode solver, which was used to calculate the post-disruption Alfv\'en continuum, shown in figure~\ref{fig:GAE}.
    In~(a) and~(b) the results from LIGKA for a fixed toroidal mode number of \(n = 0\) are presented for different time points. This mode number was chosen based on the results of the mode number analysis (section~\ref{sec:observations}).
    The plots show the coupled Alfv\'en continuum as solid lines. The shaded areas around the central lines show the results for a 20\% in-/decrease in ion density. In addition, the first four poloidal mode numbers are labeled in the plot. The ASTRA simulations yield, that the electron temperature and \(q\) profiles are highly peaked due to pre-injection application of strong central ECRH. Consequently, extrema are visible in the Alfv\'en continuum.
    
    For typical experimental setups, this extremum is most likely a minimum in the Alfv\'en continuum~\cite{Heidbrink_2008, Appert1982}.
    This is due to higher peaked plasma current profiles compared to density profiles for present day tokamaks~\cite{Appert1982}. Therefore, GAEs are expected just below the minima in the Alfv\'en continuum for typical scenarios~\cite{Heidbrink_2008, Appert1982}.
    However, during disruptions -- especially during ones mitigated via impurity injection -- the plasma profiles vary significantly. This also affects the \(q\)-profile, the Alfv\'en velocity and continuum. Higher plasma densities towards the edge region can also lead to the formation of a maximum in the continuum, which is visible in the plots in figure~\ref{fig:GAE}(a) and~(b).
    GAEs are expected to be present slightly above this maximum for these cases.
    
\begin{figure}[h]
    	\centering
    		\subfigure[LIGKA scans above \(m = 2\) maximum.]{\includegraphics[width = 0.9\linewidth]{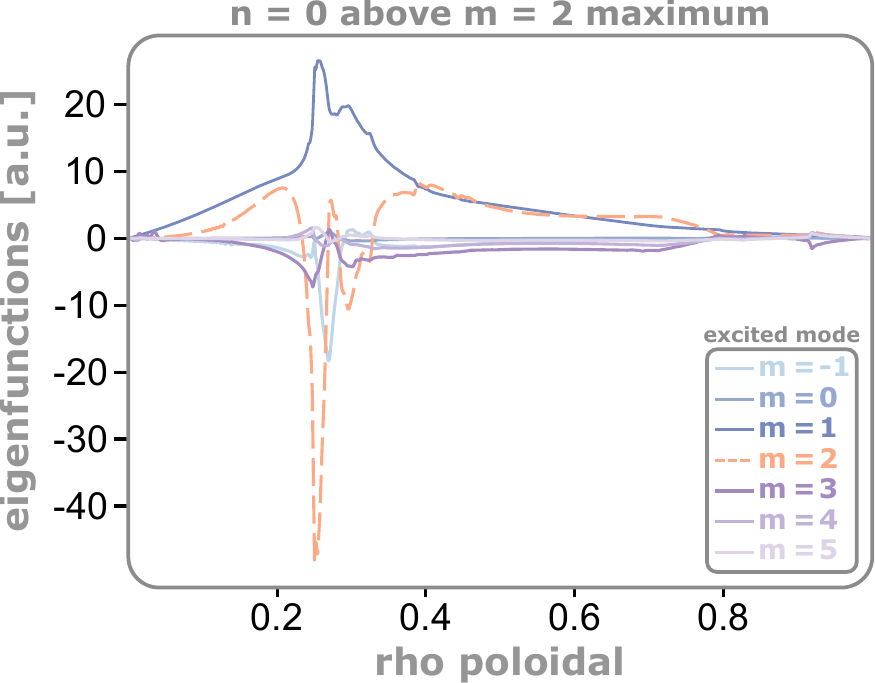}\label{fig:EF m=2}}
    		\hfil
    		\subfigure[LIGKA scans above \(m = 3\) maximum.]{\includegraphics[width = 0.9\linewidth]{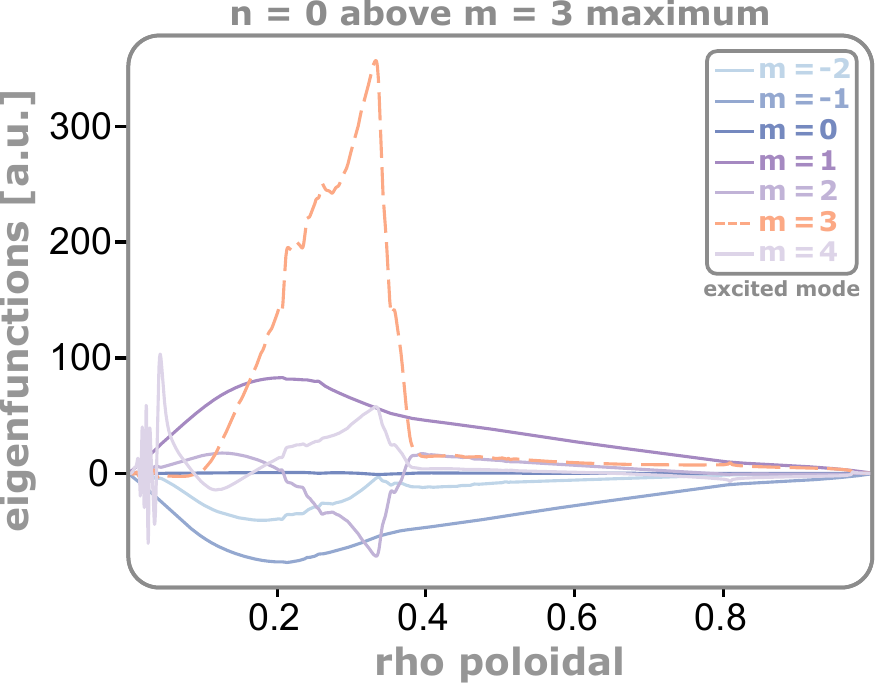}\label{fig: EF m=3}}
    	\caption{Plasma response scans in the LIGKA simulations. An artificial test-antenna is mimicked, launching waves into the plasma and monitoring the plasma response. The respective \(m = 2, 3\) contributions are highlighted by the dashed lines.}
    	\label{fig:EF plots}
    \end{figure}

    For this reason, we performed plasma response scans at these maxima (at the position of the gray arrows in figure~\ref{fig:n0_1005}) in the LIGKA simulations.
    Hereby, excitations by an artificial test-antenna are mimicked and the plasma response to these waves is monitored. Strong reactions of the plasma to these waves hint towards potential modes excited by these waves. The results of the scans above the \(m = 2\) or \(m = 3\) maximum, are given in figure~\ref{fig:EF plots}. Here the value of the eigenfunctions are plotted against the radial position and the lines represent the different mode numbers of the plasma response. The response is strongest for the mode number affiliated with the maxima of the Alfv\'en continuum scanned. Therefore, the strongest response is measured for \(m = 2\) and \(m = 3\) above the \(m = 2\) and \(m = 3\) maximum, respectively.

    As shown in figure~\ref{fig:EF plots}, we are likely to excite modes (namely GAEs) above these maxima.
    We can now use this information and track the mode behaviour for different time points shown in figure~\ref{fig:GAE}. For \(t = 1.005\ \sec\) (figure~\ref{fig:n0_1005}), we highlighted the modes with \(m = 2\) and \(m = 3\), which are in a similar frequency range as the mode observed for this discharge. The gray horizontal lines show the potential radial extent of excited modes. At the intersection with the continuum lines, continuum damping will occur, thus is very likely to set a limit to the radial extent. As the mode frequency will be close to the maxima of the peaked profiles, it is also possible to extract the frequency evolution from the plot. The frequencies, including the error bars from the 20\% change in ion density, are plotted on top of the spectrogram for the magnetic pick-up coil \mbox{B31-03} in figure~\ref{fig:theory_spec}. With time, the profiles evolve and the frequencies of the excited waves change, as indicated by the red arrows in figure~\ref{fig:n0_1008}. Following the maxima of the profiles, the mode frequency decreases as observed also in the experiment.
    At \(t = 1.008\ \sec\) (figure~\ref{fig:n0_1008}) the peaked profiles flatten and the maxima vanish. With this, we also expect the mode to cease, matching the experimental observations displayed in the spectrogram~\ref{fig:theory_spec}. However, the frequency evolutions predicted by the simulation results do not quite match the experimental results. While the \(m = 2\) mode would match the observations well for later times, the starting frequency is slightly underestimated.
    However, matching the experiment with the ASTRA simulations for discharge \#35618 had been challenging, thus future simulations might increase the precision of the frequency predictions. This first analysis of the CQ modes does not include investigations for possible mode driving mechanisms. Based on the also prominent \(n = -4\) contribution of the mode number investigations, the mode might rotate in the ion diamagnetic direction with a mode frequency below the ion cyclotron resonances. Therefore, the modes might be driven by drift reversed trapped electrons or by non-linear excitation mechanisms, such as wave-wave coupling processes.

\section{Summary \label{sec:Summary}}

For safe operation of next step fusion reactors, the suppression of runaway electron beams is an essential task. Making use of intrinsically or externally driven Alfv\'enic activity during the current quench phase could be an actuator for runaway electron mitigation. In this paper, we show the results from a detailed study on the behaviour and statistics of modes in post-disruption plasmas in ASDEX Upgrade and explain their frequency change by gaps in the GAE continuum.

The modes appear exclusively during the current quench and typically sweep from higher to lower frequencies in a range around 800~kHz to 300~kHz for the first harmonic with their most likely frequency around 400~kHz. The modes are globally coherent, however, are composed of many different mode numbers, thus do not show a unique mode number. The most prominent toroidal mode numbers are \(n = 0\) and \(n = -4\), where the negative sign indicates a propagation in the ion diamagnetic drift direction. The statistical analysis was performed on a database consisting of 180 disruptions aiming for runaway electron generation, which shows the presence of a mode in 179 cases.

The mode behaviour does not show strong correlations with global parameters, such as plasma current or magnetic field properties. Even though the mode length correlates with the subsequent runaway current, the mode amplitude does not. Moreover, the cause of this correlation might be the connection to the current quench duration, which sets the boundary conditions for the appearance of these modes. Modes were also observed in natural disruptions with no subsequent RE beam. The analysis of ``multi-stage" disruptions revealed, that the modes appear exclusively after the last $\textnormal{I}_\textnormal{p}$ spike occurred, indicating that the conditions of the mode appearance is more complex as simple critical thresholds in the plasma current. 

We performed dedicated ASTRA and LIGKA simulations to investigate the mode origin. We were able to reproduce a similar frequency evolution to our reference discharge \#35618. First, an ASTRA simulation was performed to access local plasma parameters such as the safety factor and temperature profile. LIGKA simulations for a fixed toroidal mode number of \(n = 0\) revealed gaps in the GAE-continuum, where mode activity can be excited. Following the changes in the local plasma parameters over time, those gaps shift towards lower frequency, which can explain the sweeping behaviour of these modes. Moreover, the flattening of the Alfv\'en continuum profiles at \(t = 1.008\)~sec will cause the mode to vanish, matching our experimental observations. Within the error bars, the absolute frequency is reasonably matched for a mode with an poloidal mode number of \(m = \pm 2\). The mode driving mechanisms were not investigated during these studies, however, potential candidates could be drift reversed trapped electrons or non-linear excitation mechanisms, like wave-wave coupling.

We conclude, that the modes observed at ASDEX Upgrade and reported on in this paper do not have sufficient amplitude to have a significant impact on runaway electron dynamics in these ASDEX Upgrade scenarios. It remains to be seen, if modes in a different frequency range, and / or higher amplitudes can be achieved in ASDEX Upgrade runaway electron scenarios.

\section*{Acknowledgements}
This work has been carried out within the framework of the EUROfusion Consortium, funded by the European Union via the Euratom Research and Training Programme (Grant Agreement No 101052200 — EUROfusion). Views and opinions expressed are however those of the author(s) only and do not necessarily reflect those of the European Union or the European Commission. Neither the European Union nor the European Commission can be held responsible for them.

\bibliographystyle{iaea_papp_natbib}
\bibliography{biblio.bib}

\pagebreak
\appendix

\section{Appendix}\label{sec:appendix}
    
\subsection{Correlation Matrix including 4 regions of interest} 

\newgeometry{left=10cm,right=0cm,top=0.5cm,bottom=0.5cm}
\thispagestyle{empty}
        
\begin{sidewaysfigure*}
	\begin{center}
                \includegraphics[width=\linewidth]{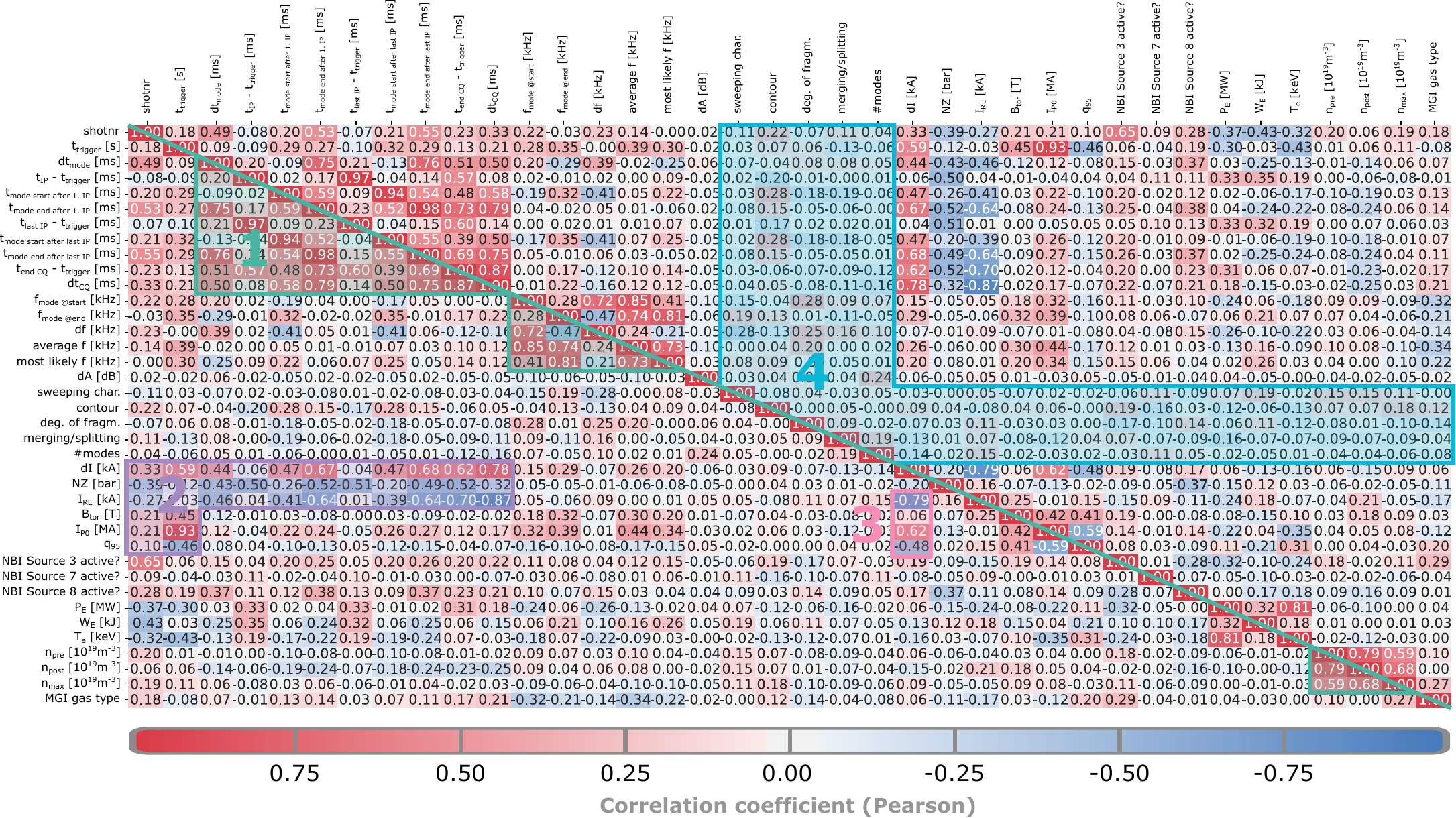}\\
                \caption{The correlation matrix with Pearson correlation coefficients.}\label{fig:CorrMatrix}
      	\end{center}
\end{sidewaysfigure*}
        
\restoregeometry
\pagebreak
    
    \begin{longtable}{|p{0.3\linewidth}|p{0.65\linewidth}|}
        \multicolumn{2}{c}{\textbf{PARAMETER DESCRIPTIONS}} \\
        \hline
        parameter [unit] & \multicolumn{1}{c|}{description}\\
        \hhline{|=|=|}
        \endfirsthead
        \hline
        parameter [unit] & \multicolumn{1}{c|}{description}\\
        \hhline{|=|=|}
        \endhead
        
        shotnr & ID of the discharge. The routine/setup might change between discharges.\\
        \hline
        \(t_\textnormal{trigger}\)~[s] & Time when the MGI valve was triggered. \mbox{\(t_\textnormal{trigger} = 1.0\) s} for 160/180 discharges.\\
        \hline
        \(\textnormal{d}t_\textnormal{mode}\)~[ms] & Duration of the mode given by the mode tracing tool (user selection of start and end points by hand).\\
        \hline
        \(t_\textnormal{IP} - t_\textnormal{trigger}\)~[ms] & Time point of the \highlight{initial} $\textnormal{I}_\textnormal{p}$ spike.\\
        \hline
        \(t_\textnormal{mode start after 1. IP}\)~[ms], \(t_\textnormal{mode end after 1. IP}\)~[ms] & Start \& end time of the mode with respect to the \highlight{initial} $\textnormal{I}_\textnormal{p}$ spike\\
        \hline
        \(t_\textnormal{last IP} - t_\textnormal{trigger}\)~[ms] & Time point of the \highlight{last} $\textnormal{I}_\textnormal{p}$ spike for multi-stage disruptions.\\
        \hline
        \(t_\textnormal{mode start after last IP}\)~[ms], \(t_\textnormal{mode end after last IP}\)~[ms] & Start \& end time of the mode with respect to the \highlight{last} $\textnormal{I}_\textnormal{p}$ spike\\
        \hline
        \(t_\textnormal{end CQ} - t_\textnormal{trigger}\)~[ms] & End time with respect to the trigger time.\\
        \hline
        \(\textnormal{d}t_\textnormal{CQ}\)~[ms] & Length of the CQ. Defined by the initial $\textnormal{I}_\textnormal{p}$ spike and when the temporal derivative approaches 0 again.\\
        \hline
        \(f_\textnormal{mode \@start}\)~[kHz] & Starting frequency of the mode.\\
        \hline
        \(f_\textnormal{mode \@end}\)~[kHz] & Frequency at the end of the mode.\\
        \hline
        d\(f\)~[kHz] & Change in frequency \(f_\textnormal{mode \@end} - f_\textnormal{mode \@start}\)\\
        \hline
        average \(f\)~[kHz] & Average mode frequency derived by the arithmetic mean.\\
        \hline
        most likely \(f\)~[kHz] & Most likely frequency of the mode (see section~\ref{subsec:MLFs}).\\
        \hline
        \(A_\textnormal{min}\)~[dB], \(A_\textnormal{max}\)~[dB] & Minimum/Maximum mode amplitude.\\
        \hline
        d\(A\)~[kHz] & Difference in mode amplitude \(A_\textnormal{max} - A_\textnormal{min}\).\\
        \hline
        sweeping char.\footnote{The characteristics are introduced in section~\ref{sec:Classification}} & Rating of the sweeping characteristic from \highlight{typical}~(1) -- \highlight{{sinusoidal}}~(5) and \highlight{no mode}~(0).\\
        \hline
        contour\footnotemark[1] & Rating of the mode contour from \highlight{narrow}~(1) -- \highlight{very cloud like}~(5) and \highlight{no mode}~(0).\\
        \hline
        deg. of fragm.\footnotemark[1] & Rating of the degree of fragmentation from \highlight{not fragmented}~(1) -- \highlight{strongly fragmented}~(5) and \highlight{no mode}~(0).\\
        \hline
        merging/splitting\footnotemark[1] & Rating of the mode merging \& splitting behaviour from \highlight{not merging nor splitting}~(1) -- \highlight{splitting and merging}~(5) and \highlight{no mode}~(0).\\
        \hline
        \#modes\footnotemark[1] & Number of prominent modes inside the CQ.\\
        \hline
        \(I_\textnormal{P0}\)~[MA] & Pre-disruption plasma current.\\
        \hline
        \(I_\textnormal{RE}\)~[kA] & Runaway current after the CQ.\\
        \hline
        d\(I\)~[kA] & Difference between the pre-disruption and runaway current.\\
        \hline
        NZ~[bar] & Amount of MGI gas injected.\\
        \hline
        \(B_\textnormal{tor}\)~[T] & Toroidal magnetic field strength.\\
        \hline
        \(q_\textnormal{95}\) & Value of the safety factor at \(\rho_\textnormal{pol} = 0.95\).\\
        \hline
        NBI-source 3/7/8 active? & States whether the respective NBI source 3, 7 and 8 is active (0~=~not active,~1~=~active).\\
        \hline
        \(P_\textnormal{E}\)~[MW] & Average heating power by ECRH before the valve is triggered.\\
        \hline
        \(W_\textnormal{E}\)~[kJ] & Plasma stored energy.\\
        \hline
        \(T_\textnormal{e}\)~[keV] & Electron temperature measured by ECE-measurements.\\
        \hline
        \(n_\textnormal{pre}\)~[\(10^{19}~\textnormal{m}^{-3}\)] & Pre-disruption plasma density measured by \(\textnormal{CO}_2\) interferometry (LOS: V-1).\\
        \hline
        \(n_\textnormal{post}\)~[\(10^{19}~\textnormal{m}^{-3}\)] & Post-disruption plasma density measured by \(\textnormal{CO}_2\) interferometry (LOS: V-1).\\
        \hline
        \(n_\textnormal{max}\)~[\(10^{19}~\textnormal{m}^{-3}\)] & Maximum plasma density measured by \(\textnormal{CO}_2\) interferometry (LOS: V-1).\\
        \hline
        MGI gas type & One of the following four gas types are used for the MGI: Ar, Ne, Kr or Xe with numbers 1--4 assigned.\\
        \hline
        \caption{Excerpt of parameters inside the database.}
        \label{tab:CM parameters}
    \end{longtable}

\subsection{Highlighted regions in the correlation matrix}~\label{sec:CorrMat_regions}

\textbf{Region 1} shows high (positive) correlation values, thus is one of the first regions drawing the attention onto it. It is attached to the principal diagonal, therefore it shows the correlation results for factors close to each other in the factor list. It depicts the three blocks of time-, frequency- and density-components. Let us pick the frequency block as an example. If we have a high starting frequency (\(f_\textnormal{mode @start}\) [kHz]), the difference in frequency (d\(f\) [kHz]) and also the average frequency (average \(f\) [kHz]) is more likely to be high. As a result, their correlation value is close to +1. Similar internal connections also apply to the other blocks of region 1.\\
        
For \textbf{region 2} this is not the case anymore. It is located far from the principal diagonal and connects the \highlight{time block} with \highlight{global parameters} such as the (runaway) current, toroidal magnetic field strength and \(q_{95}\). Let us focus on the current measurements first.
The mode lives inside the CQ as indicated in figure~\ref{fig:CQ_location}. For longer current quench and mode duration, lower subsequent currents are held by runaway electrons (\(I_\textnormal{RE}\) [kA]) and a higher change in current (d\(I\) [kA]) will be achieved (see figure~\ref{fig:I_RE plots}). This results in high correlation values -- positive for the change in current, negative for the runaway current. Another observation of this block is the anti-correlation between the amount of gas injected to trigger the disruption ($\textnormal{N}_\textnormal{Z}$ [bar]) and the start (\(t_\textnormal{IP} - t_\textnormal{trigger}\)~[ms]) and end (\(t_\textnormal{end CQ} - t_\textnormal{trigger}\) [ms]) times of the current quench, respectively. The values for the safety factor \(q_{95}\) are typically adjusted by varying the plasma current and toroidal magnetic field. As the pre-disruption plasma current is determined by the timing of the MGI, the trigger time shows correlations with \(B_\textnormal{tor}\) and \(q_{95}\). Thereby, earlier injections lead to lower plasma currents, hence higher safety factors.\\
        
In \textbf{region 3} the correlation between the change in current (d\(I\)~[kA]) and respective runaway current (\(I_\textnormal{RE}\)~[kA]), pre-disruption current (\(I_\textnormal{P0}\) [MA]) and q95, are shown. It is expected for d\(I\) and \(I_\textnormal{RE}\) to anti-correlate, as well as d\(I\) and \(I_\textnormal{P0}\) to correlate, as already indicated in the explanations for region~2. The correlation between \(q_{95}\) and d\(I\), can be linked to two different factors. First again \(q_{95}\) and \(I_\textnormal{P0}\) are connected as the plasma current is one of the factors determining \(q\). Hence, lower plasma currents often result in higher safety factors and at the same time only allow reduced changes in plasma current. Additionally, it is observed for ASDEX Upgrade, that if \(q_{95} < 3\) no RE plateau is forming~\cite{Papp_EPS_2019}. As no RE plateau is present for safety factors below three, the potential change in current is higher.\\
        
\textbf{Region 4} shows the correlation values of the mode characteristics introduced in section~\ref{sec:Classification}. This region is of special interest, as high correlation results would point towards a connection between the mode behaviour and global parameters. However, we do \highlight{not} observe any significant correlations in this region! The higher values within this region, e.g. for the correlation of the change in frequency (d\(f\)~[kHz]) with the sweeping characteristic, can be explained due to the rating system. For this example the sweeping characteristic is granted a number between 1~(\textit{typical}) to 5~(\textit{sinusoidal}) (and 0 for \textit{no mode}). Number 3 is assigned to the behaviour \textit{not chirping}. The anti-correlation is caused by the \textit{typical} sweeping behaviour having a larger change in frequency compared to the \textit{not chirping} one. So by increasing the rating-number from 1 to 3 the change in frequency decreases. Similar effects of the rating system are observed for other characteristics.

\subsection{Potential explanations for the loadings of each principal component in figure~\ref{fig:FA}}
\begin{enumerate}\label{enum:FA_descriptions}
	\item[1.] The first underlying factor~\texttt{1} groups mainly the end times of the modes with the current evolution, which was discussed for region 2 in section~\ref{subsec:CorrMat2}.
        \item[2.] Here, the frequency properties of the modes are addressed. In section~\ref{subsec:MLFs} a comparison between the average and most likely frequency is provided. Hereby, the end frequency of the mode has the most influence on the average and most likely frequency, respectively. Therefore, the loading for the start frequency is lower compared to the other loadings.
        \item[3.]This factor illustrates the current ramp-up phase. When the MGI valve is triggered at later times, higher pre-disruption plasma currents and lower safety factors are present. This also affects the possible change in current d\(I\).
        \item[4.] Higher amounts of gas injected into the chamber will lead to earlier $\textnormal{I}_\textnormal{p}$ spike times and ultimately also the end time of the current quench reduces.
        \item[5.] In this principal component, the frequency properties of the mode are addressed. The start frequency and change in frequency show the highest loadings. The change in frequency is calculated via both -- start and end frequency -- however, the variation of the start frequency is typically larger compared to the range of end frequencies.
        Smaller loadings for the sweeping behavior and degree of fragmentation and are also visible. A possible explanation for this is presented in the following, with discharge \#35624 in figure~\ref{fig:Spectrograms}(c) as an example. As the beginning of the mode does not show a continuous mode evolution, it was rated as ''strongly fragmented''. The starting frequency of the mode is around 680~kHz and the other part of the mode between 400 -- 370~kHz. In case of reduced visibility of the mode in the spectrogram -- due to a bad SNR or any physical reason -- the traced mode would start at later times with a lower frequency. Accordingly, a decreased change in frequency and mode duration would be measured. In addition, without the first fragments of the mode being visible, this mode would not be considered as ''fragmented'' anymore, hence, the degree of fragmentation would decrease, which could explain the loadings in this principal component. In-/excluding the first part of the mode for the discharge \#36425 would not affected the MLF, as it is mainly determined by the end frequency of the mode at the flat tail. This matches the MLFs-loading close to zero.
        \item[6.] The plasma density measurements at different times are connected via factor~\texttt{6}. For higher pre-disruption densities also a higher post-disruption and maximum density is expected.
        \item[7.] Here, mainly the start times of the modes after the first and last $\textnormal{I}_\textnormal{p}$ spike are connected. As only five discharges within the database contain multiple $\textnormal{I}_\textnormal{p}$ spikes during incomplete disruptions (see figure~\ref{fig:incomplete_disruption_35636}), mostly the first $\textnormal{I}_\textnormal{p}$ spike is also the last $\textnormal{I}_\textnormal{p}$ spike at the same time. Therefore, the values match most of the time. Additionally, smaller loadings for the change in frequency and mode contour characteristic are contained. For this, two effects could explain these loadings:
        Firstly, the accuracy of the correct start time of the mode. For larger mode contour rating numbers -- especially the "extended" modes, it is increasingly difficult to pin down an exact start and end time of the mode. Therefore, the actual start of the mode might be sometimes earlier than determined by hand. A more detailed illustration of this is given in the thesis by P. Heinrich~\cite{Heinrich_2021_MSc}. As the modes often chirp strongest at their start, the change in frequency is affected by this as well. 
        On the other hand, extended modes were observed have a lower chirping rate on average compared to modes with a smaller spectral width.
        \item[8.] Illustration of changes in the heating schemes over the course of the experimental campaign. In later discharges, the NBI source 3 with different heating schemes were used compared to the older discharges.
        \item[9.] The safety factor is directly proportional to the strength of the toroidal magnetic field component, which is illustrated by the loadings for factor~\texttt{9}.
\end{enumerate}

\end{document}